\documentclass[fleqn,10pt,twocolumn]{wlscirep} 

\usepackage[utf8]{inputenc}
\usepackage[T1]{fontenc}
\usepackage{tablefootnote}
\usepackage{array}
\usepackage[quotient-mode = fraction]{siunitx}
\usepackage{todonotes}

\usepackage{enumitem} 
\usepackage{algorithm}
\floatname{algorithm}{Procedure}
\usepackage{soul} 

\title{Machine-learning potentials for nanoscale simulations of deformation and fracture: example of TiB$_2$ ceramic}

\author[1,2*]{Shuyao Lin}
\author[2]{Luis Casillas-Trujillo} 
\author[2]{Ferenc Tasnádi}
\author[2]{Lars Hultman}
\author[1]{Paul H. Mayrhofer}
\author[2]{Davide G. Sangiovanni}
\author[1,2]{Nikola Koutná}
\affil[1]{Technischen Universität Wien, Institute of Materials Science and Technology, Vienna, A-1060, Austria}
\affil[2]{Linköping University, Department of Physics, Chemistry, and Biology (IFM), Linköping, SE-58183, Sweden}

\affil[*]{shuyao.lin@tuwien.ac.at}


\begin{abstract}
Machine-learning interatomic potentials (MLIPs) offer a powerful avenue for simulations beyond length and timescales of {\it{ab initio}} methods. 
Their development for investigation of mechanical properties and fracture, however, is far from trivial since extended defects---governing plasticity and crack nucleation in most materials---are too large to be included in the training set. 
Using TiB$_2$ as a model ceramic material, we propose a strategy for fitting MLIPs suitable to simulate mechanical response of monocrystals until fracture.
Our MLIP accurately reproduces {\it{ab inito}} stresses and failure mechanisms during room-temperature uniaxial tensile deformation of TiB$_2$ at the atomic scale ($\approx{10}^3$ atoms).
More realistic tensile tests (low strain rate, Poisson's contraction) at the nanoscale ($\approx{10}^4$--10$^6$ atoms) require MLIP up-fitting, i.e. learning from additional {\it{ab initio}} configurations. 
Consequently, we elucidate trends in theoretical strength, toughness, and crack initiation patterns under different loading directions.
To identify useful environments for further up-fitting, i.e., making the MLIP applicable to a wider spectrum of simulations, we asses transferability to other deformation conditions and phases not explicitly trained on.

\end{abstract}
\begin{document}
\flushbottom
\maketitle
\thispagestyle{empty}

\section*{Introduction}
Simulations of materials' mechanical response---including (i) intrinsic strength and toughness, (ii) nucleation of extended defects (e.g. dislocations, stacking faults, cracks) and their implications for (iii) plasticity and fracture mechanisms~\cite{liu2013stacking,sabzi2016effects,yang2017impact}---require length and time scales beyond limits of {\it{ab initio}} methods ($\approx{10^3}$ atoms, $\ll$ns)\cite{khosrownejad2021quantitative,bianchini2019enabling,zhao2023understanding}.
The {\it{go-to}} method in most cases would be Molecular Dynamics (MD), allowing to access atomistic pathways for deformation and fracture in nanoscale systems ($\approx{10^6}$ atoms) and ``realistic'' operation conditions (e.g. ultra-high temperatures, times up to $\mu$s).
However, a severe problem of classical MD is that the necessary interatomic potentials do not exist for most engineering materials, or, are limited in accuracy and transferability with respect to temperatures, phases, and defect structures (see e.g. Refs.~\cite{fail2,fail1,proville2012quantum}).

A powerful avenue for MD simulations on multiple time and length scales with near {\it{ab initio}} accuracy is the application of machine learning interatomic potentials~\cite{deringer2019machine,zuo2020performance,podryabinkin2019accelerating,add1} (MLIPs, in case of no ambiguity just ``potentials'').
MLIPs learn the atomic energy (or other atomic properties) as a nontrivial function of a descriptor quantifying local atomic environments in an {\it{ab initio}} training set\cite{drautz2019atomic}.
Compared to conventional {\it{ab initio}} calculations, MLIPs can achieve a speed up of as much as 5 orders of magnitude\cite{smith2017ani,shapeev2020elinvar}. 
Previous studies showed examples of MLIPs' transferability with respect to defects (e.g. grain boundaries~\cite{add2}, dislocation structures~\cite{deng2023large,mori2020neural}) and phases~\cite{rosenbrock2021machine,zong2018developing} (e.g. Ni-Mo phase diagram illustrating superior performance of a MLIP over a classical potential~\cite{fail3}).
Recently, Tasn\'{a}di et al.~\cite{Method4} have demonstrated high accuracy of MLIP-predicted elastic constants for TiAlN ceramics, hence, have set the stage for MLIP development beyond linear elastic regime.

Based on the parametrization of local structural properties, MLIPs can be fitted employing different formalisms: spectral neighbor analysis potentials (SNAP)~\cite{thompson2015spectral}, neural networks potentials (NNP)~\cite{behler2011atom,behler2015constructing}, Gaussian approximation potentials (GAP)~\cite{bartok2010gaussian}, moment tensor potentials (MTP)~\cite{MTP}, linearized interatomic potentials~\cite{seko2019group}, or atom cluster expansion (ACE) potentials~\cite{drautz2019atomic}, where only the last three are mathematically {\it{complete}}~\cite{PACE}.
Though some parametrizations have been formally expressed as a special case of ACE~\cite{PACE}, their particular implementations may be more suitable for different training datasets and applications.
Benchmarks for competing parametrizations have been published in case of carbon~\cite{qamar2022atomic} or graphene~\cite{rowe2018development}, but are missing for chemically complex materials.

Besides various MLIP formalisms, additional fundamental challenges in the field are: (i) efficient training dataset generation (i.e. minimising computationally-intensive {\it{ab initio}} calculations), (ii) standardized validation procedure, and (iii) training strategies for simulations beyond {\it{ab initio}} reach. 
A solution to (iii) would facilitate nanoscale description of solids with atomic-scale resolution, hence, is the key to understanding differences between  material's sub-micrometer and macroscopic behavior~\cite{nie2020direct,sivaraman2020machine,fiedler2023predicting}. 
Point (iii) is also a primary motivation of this study which presents methodological (MLIP development) as well as materials science discussion of nanoscale MD simulations on deformation and fracture for a model TiB$_2$ ceramic.

The chosen material, TiB$_2$, is a widely researched representative of ultra-high temperature ceramics (UHTCs) with mature synthesis technologies~\cite{chen2021room,zhang2019preparation} and melting point of 3500~K~\cite{munro2000material}.
UHTCs exhibit high hardness and resistance to oxidation, corrosion, abrasive and erosive wear~\cite{magnuson2022review,holleck1986material}, thus, are suitable to protect tools and machining components under extreme conditions~\cite{golla2020review,wang1995electrical,sevik2022high,wiedemann1997structure}.
Similar to other transition metal diborides, TiB$_2$ crystallizes in the hexagonal $\mathrm{AlB}_2$-type phase~\cite{hofmann1936struktur,eorgan1967zirconium,norton1949structure} ($\alpha$, P6/mmm) and shows outstanding mechanical properties~\cite{mikula2007mechanical,geng2017microstructural} including hardness of 41--53~GPa\cite{H1,H2,H3}.
From computational materials science perspective, insights into mechanical behaviour of TiB$_2$ and other diborides have been offered by {\it{ab initio}} calculations~\cite{zhou2014first,dai2016effects,zhang2010ideal} and recently also by molecular dynamics with classical empirical potentials (TiB$_2$\cite{attarian2022development,attarian2021multiscale}, ZrB$_2${\cite{timalsina2021development}, HfB$_2$\cite{daw2011interatomic}). 
To date, no MLIPs capable of predicting mechanical response of UTHCs until fracture, however, have been reported.

Here, we propose a training strategy for MLIPs suitable to model deformation and crack growth mechanisms in monocrystals from atomic to nanoscale.
Viability of our approach is illustrated by MD simulations of room-temperature uniaxial tensile loading for supercells with $\approx{10}^3$--10$^6$ atoms.
Specifically, we use the MTP formalism 
and train on a small fraction (initially $<1\%$) of snapshots from room-temperature {\it{ab initio}} molecular dynamics (AIMD) tensile tests, iteratively expanding the training set until all configurations are reliably described. 
Our MLIP accurately reproduces stress evolution and failure mechanisms during tensile loading of TiB$_2$ at the atomic scale, but requires up-fitting for more realistic nanoscale tensile tests.
Consequently, we discuss size effects in the predicted mechanical properties and fracture patterns, as well as asses MLIP's transferability with respect to other loading conditions and phases.

\section*{Results and discussion}
\subsection*{1 Training procedure and fitting initial MLIPs}
Our general training procedure is described below (Procedure~\ref{training procedure}) and schematically depicted in Fig.~\ref{FIG: workflow}a.
Throughout this work, {\it{ab initio}} training data is generated by finite-temperature AIMD calculations generally producing many highly correlated configurations\footnote{By a configuration, we mean a structure labelled by {\it{ab initio}} total energy, forces acting on each ion, and six stress tensor components.}. 
To avoid over-representation, our MLIP training initiates with a small fraction of AIMD configurations and exploits the MLIP's uncertainty indication---quantified through the extrapolation/Maxvol grade\cite{MV} ($\gamma$)---to iteratively expand the training set until all AIMD configurations are reliably described.
Mathematically grounded in Refs.~\cite{MV,gubaev2018machine,lysogorskiy2023active}, $\gamma$ of a given configuration expresses how much MLIP extrapolates when predicting the corresponding energy, forces, and stresses.
Specifically, $\gamma\leq1$ means interpolation and $\gamma>1$ extrapolation. 

Note that some degree of extrapolation is acceptable.
Following Shapeev and co-workers~\cite{podryabinkin2023mlip} who describe $\gamma\leq2$ as {\it{accurate}} extrapolation, we employ a $\gamma$ threshold, $\gamma_{\text{thr}}=2$, as a condition for exiting the training loop in Procedure~\ref{training procedure}. 

\begin{algorithm}[!htbp]
\caption{MLIP training}\label{training procedure}
\begin{itemize} \setlength\itemsep{0em} 
    \item[(1)] Generate a pool of AIMD configurations.
    
    \item[(2)] Divide the pool into an initial training set (TS$_0$), a learning set (LS), and a validation set (VS) by randomly selecting 0.5\%, 79.5\%, and 20\% of non-overlapping configurations.
    
    \item[(3)] Fit an initial MLIP (MLIP$_0$, trained on TS$_0$). If $\gamma$ of all configuration in the LS and VS is below $\gamma_{\text{thr}}=2$, exit. Else, build TS$_1$ by adding (maximum 15) highly extrapolative configurations from the LS to TS$_0$ and fit a new MLIP (MLIP$_1$, trained on TS$_1$). 
    
    \item[(4)] While $\gamma$ of all configurations in the LS and VS is above $\gamma_{\text{thr}}=2$, build TS$_i$ by adding (maximum 15) highly extrapolative configurations from the LS to TS$_{i-1}$, and fit a new MLIP (MLIP$_{i}$, trained on TS$_i$).
\end{itemize}
\end{algorithm}

\newpage  
Technical comments on Procedure~\ref{training procedure} (for further details of MLIP training, please see the Methodology):
\begin{itemize} \setlength\itemsep{0em} 
    \item MLIP$_0$ in the step (3) is trained from an untrained MTP. To speed-up the fitting process, MLIP$_i$ in step (4) is fitted from MLIP$_{i-1}$ if maximum $\gamma$ in the $(i-1)$th iteration is below 1000 (otherwise it is fitted from an untrained MTP).
    
    \item The VS is not used for training but only as a reference\footnote{Although it was not the case here, the maximum $\gamma$ of VS may remain high ($\gamma\gg\gamma_{\text{thr}}$) even if the maximum $\gamma$ of LS is already below $\gamma_{\text{thr}}$ (imagine, e.g., all configurations relevant for the description of material's fracture ending up in the VS). Then the strategy could be switching the LS and VS, i.e. use the old VS for learning and the old LS for validation.}.
    
    \item Besides $\gamma$, quality of the fit at each iteration $i$ in (4) is monitored through errors of energies, forces and stresses (quantified by common regression model evaluation metrics, MAE, RMSE, R$^2$, see e.g. Refs.~\cite{chicco2021coefficient,willmott1982some,figueiredo2011r2}) for the TS$_i$ (fitting errors) and the VS (validation errors)\footnote{While not done here, one may use fitting and validation errors as additional criterion (besides $\gamma_{\text{thr}}$) for exiting the training loop.}. 
\end{itemize}

Employing Procedure~\ref{training procedure} and the MTP formalism, we fit three MLIPs: MLIP-[0001], MLIP-[10$\overline{1}0$], and MLIP-[$\overline{1}2\overline{1}0$].
The training uses snapshots from room-temperature AIMD simulations for a 720-atom TiB$_2$ supercell, uniaxially elongated in the [0001], $[10\overline{1}0]$, and $[\overline{1}2\overline{1}0]$ crystallographic direction, respectively, with a strain step of 2\% (for details of AIMD simulations, see the Methodology).
The entire pool of AIMD data consists of $\approx{120,000}$ configurations, where each loading condition ([0001], $[10\overline{1}0]$, and $[\overline{1}2\overline{1}0]$) represents $\approx{1/3}$\footnote{Since we run AIMD tensile tests until fracture, directions with higher fracture strains produce more configurations.}. 

The final training sets (the last TS$_i$ in the step (4) of  Procedure~\ref{training procedure}) of MLIP-[0001], MLIP-[10$\overline{1}0$], and MLIP-[$\overline{1}2\overline{1}0$] contain 181, 155, and 180 configurations, respectively.
The fitting and validation errors, quantified through the residual mean square error (RMSE), of total energies, forces, and stresses do not exceed 2.6\;meV/at., 0.11\;eV/\AA, and 0.19\;GPa, respectively.

\begin{figure*}[!htbp]
\centering
\includegraphics[width=1.02\textwidth]{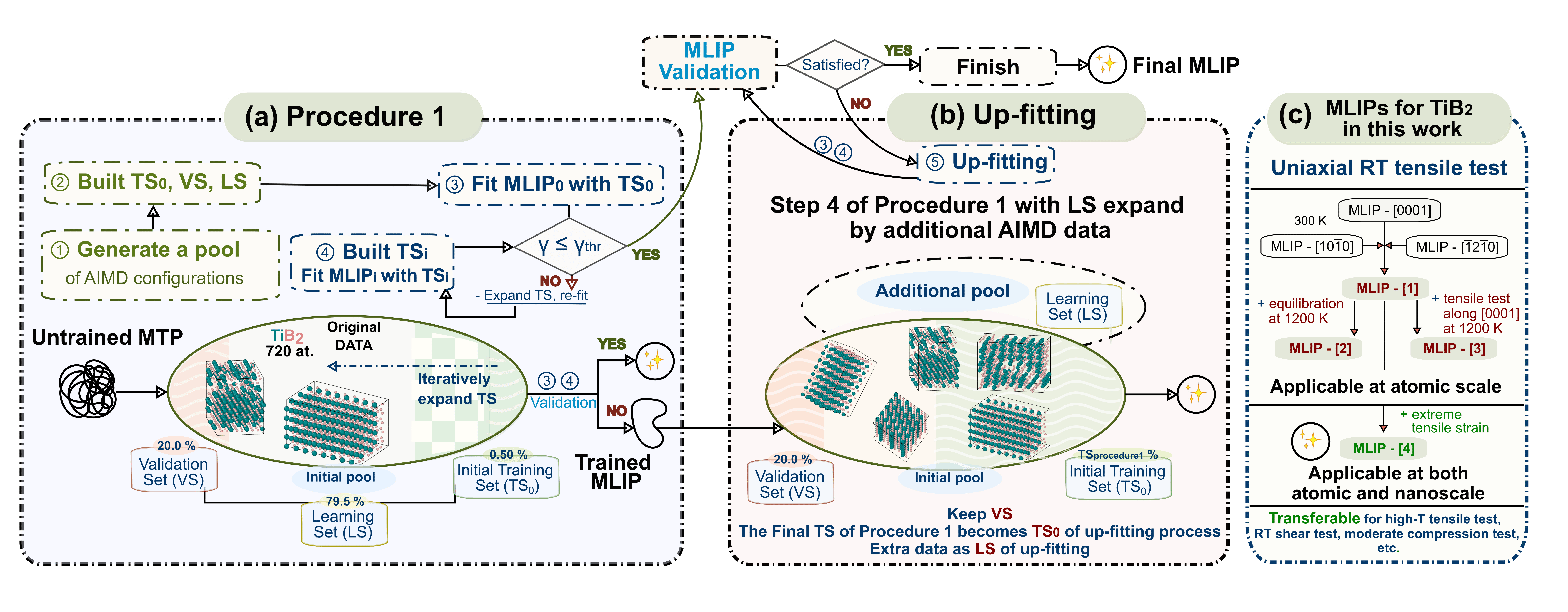}
\caption{\footnotesize
Schematic visualization of (a) our general training procedure (Procedure~\ref{training procedure}, Section 1), and (b) up-fitting (Section 2, 3).
(c) Relationships between MLIPs developed in this work (Section 1, 2, and 3) showing how MLIP-[4] was produced by up-fitting, starting from MLIP-[0001].
    }
\label{FIG: workflow}
\end{figure*}

\subsection*{2 MLIPs' validation against atomic scale tensile tests}
As a next step, the above developed MLIPs are employed for MD calculations, from now on referred to as ML-MD.
Specifically, we simulate uniaxial tensile deformation of TiB$_2$ with computational setup equivalent to that used in AIMD.
The aim is two-fold:
\begin{itemize}\setlength\itemsep{0em} 
    \item[(i)] {\ul{To asses accuracy of quantities relevant for the targeted MLIP's application and possibly measurable in experiment}}. Here, key descriptors of mechanical response are (time-averaged) stresses in the loaded direction and quantities characterizing material's strength and toughness. 
    
    \item[(ii)] {\ul{To decide if MLIP up-fitting is necessary}}.
    By up-fitting we understand expanding the LS by additional {\it{ab initio}} configurations and going back to the step (4) of Procedure~\ref{training procedure} (see Fig.~\ref{FIG: workflow}b,c).
    Up-fitting is triggered when $\gamma$ during a ML-MD simulation exceeds {\it{reliable}} extrapolation\footnote{From our training procedure (Procedure \ref{training procedure}), it follows that the final MLIP yileds $\gamma\leq\gamma_{\text{thr}}=2$ for all {\it{ab initio}} configurations in the TS, VS, and LS. Finite-temperature effects, however, may cause $\gamma>\gamma_{\text{thr}}$, even during a ML-MD simulation with computational setup equivalent to that used to generate the training data (since ML-MD trajectories will not be exactly the same as those in AIMD).
    Then, (some of) the configurations causing $\gamma>\gamma_{\text{reliable}}$ can be simply labelled by {\it{ab initio}} energies, forces, stresses and added to the LS for up-fitting (note that this is a typical definition of {\it{active learning}}~\cite{cohn1996active}). 
    At scales where direct {\it{ab initio}} calculations are unfeasible, additional configurations for up-fitting need to be generated in a different way, example of which is given in the following section. }.
    Following Ref.~\cite{podryabinkin2023mlip} we use $\gamma_{\text{reliable}}\leq10$ and show that such choice provides a good accuracy of stresses during atomic scale ML-MD tensile tests in relation to AIMD results.
\end{itemize}

Fig.~\ref{FIG: stress/strain small}a depicts stress/strain curves derived from room-temperature AIMD and ML-MD tensile tests, in which TiB$_2$ supercell ($\approx{10^3}$ atoms, $\approx{1.5^3}$~nm$^3$) is loaded in the $[0001]$, $[10\overline{1}0]$, and $[\overline{1}2\overline{1}0]$ direction, respectively. 
Note that each deformation is simulated with a MLIP trained to the respective loading condition (e.g., MLIP-[0001] for the [0001] tensile test etc.). 
Excellent quantitative agreement between AIMD and ML-MD results indicates reliability of our MLIPs.
Specifically, (time-averaged) stresses in ML-MD differ from AIMD values by 0.07--1.94~GPa, yielding statistical errors RMSE~$\approx{1.02}$~GPa, R$^{2}$~$\approx{0.9997}$\footnote{Note that stresses normal to the loaded direction---not vanishing due to the omission of Poisson's effect in both AIMD and ML-MD simulations---are also used for quantitative comparison.}.
The fracture point in [0001] deformation is excluded from the analysis (there, the [0001] stress component in AIMD does not drop to zero due to long-range electrostatic effects absent in ML-MD).
The extrapolation grade during all ML-MD simulations remains low ($\gamma\leq5<\gamma_{\text{reliable}}$), thus, suggests reliable extrapolation and does not trigger MLIP up-fitting.


\begin{figure*}[!htbp]
\centering
\includegraphics[width=0.96\textwidth]{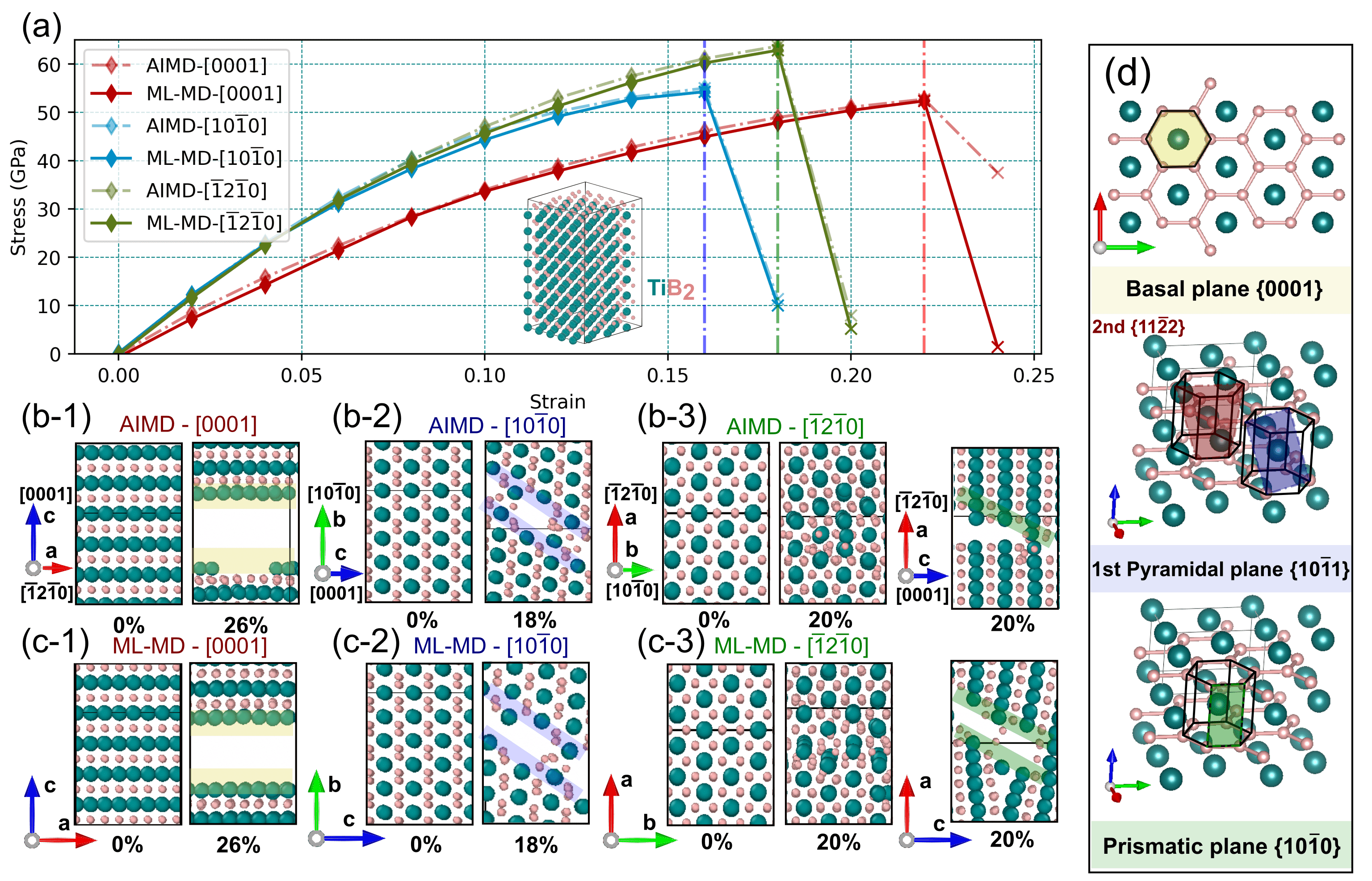}
\caption{\footnotesize
    Validation of the here-developed MLIPs (MLIP-[0001], MLIP-[10$\overline{1}0$], and MLIP-[$\overline{1}2\overline{1}0$]).
    (a) Comparison of AIMD and ML-MD stress/strain curves for TiB$_2$ subject to $[0001]$, $[10\overline{1}0]$, and $[\overline{1}2\overline{1}0]$ tensile deformation at 300~K, using a 720-atom supercell with dimensions of $\approx{(1.52\times 1.58\times 2.57)\;\mathrm{nm}^{3}}$. 
    Only the stress component in the loaded direction is plotted.
    (b, c) Snapshots of the fracture point in AIMD (b-1, b-2, b-3) and ML-MD (c-1, c-2, c-3). 
    (d) Terminology used for fracture surfaces (following Ref.~\cite{fan2015molecular}).
    }
\label{FIG: stress/strain small}
\end{figure*}

Furthermore, we use the predicted stress/strain data to evaluate TiB$_2$'s (theoretical, intrinsic) tensile strength and toughness in the $[0001]$, $[10\overline{1}0]$, and $[\overline{1}2\overline{1}0]$ direction.
By tensile strength we mean the {\it{ultimate}} tensile strength, which corresponds to the global stress maximum during the tensile test and may be reached at a strain beyond the yield point\cite{Method3}.
By tensile toughness (or toughness modulus) we understand the integrated stress/strain area until the fracture point. This property describes the ability of a material with no initial cracks to absorb mechanical energy until failure.
The $[\overline{1}2\overline{1}0]$, $[10\overline{1}0]$, and $[0001]$ tensile strength in ML-MD reaches 63.7, 55.0, and 52.7\;GPa, respectively, differing from AIMD values 
by maximum 0.8\;GPa (0.99\%). 
The ML-MD predicted toughness along $[\overline{1}2\overline{1}0]$, $[10\overline{1}0]$, and $[0001]$ 
reaches 4.3, 3.1, and 4.3\;GPa, respectively, differing from AIMD values by maximum 0.029\;GPa (0.67\%). 
Note, however, that theoretical tensile strength and toughness values may be size-dependent and should saturate at large-enough supercell sizes (see nanoscale simulations in the following sections).

For MLIPs targeted to simulations of deformation and fracture, additional aspect of validation is qualitative analysis of strain-induced structural changes.
This means confirming that the MLIP captures phenomena as, e.g., transformation toughening mechanisms or material's cleavage predicted by equivalent AIMD simulations. 
Here, our MLIPs correctly reproduce brittle failure pathways observed in AIMD of TiB$_2$ subject to [0001], $[10\overline{1}0]$, and $[\overline{1}2\overline{1}0]$ tensile deformation (Fig.~\ref{FIG: stress/strain small}b--c), as also indicated by the agreement between ML-MD and AIMD stress/strain curves (Fig.~\ref{FIG: stress/strain small}a).
The TiB$_2$'s fracture strains (same in AIMD and ML-MD) are 24\%, 20\%, and 18\% during [0001], $[10\overline{1}0]$, and $[\overline{1}2\overline{1}0]$ tensile tests, respectively.

Snapshots of the simulation cell (Fig.~\ref{FIG: stress/strain small}b--c) reveal formation of fracture surfaces with orientation equivalent to that in AIMD. 
Specifically, fracture surface for [0001] deformation almost perfectly aligns with (0001) basal planes (Fig.~\ref{FIG: stress/strain small}-(b-1),(c-1),d).
Consistently with AIMD, ML-MD tensile loading along the$[10\overline{1}0]$ direction opens a void diagonally across Ti/B$_{2}$ layers (Fig.~\ref{FIG: stress/strain small}-(b-2),(c-2)).
The fracture surface is parallel to the second order pyramidal planes of the $\{11\overline{2}2\}$ family (Fig.~\ref{FIG: stress/strain small}d). 
For the $[\overline{1}2\overline{1}0]$ loading condition, fracture planes are approximately parallel to the \{10$\overline{1}0\}$ prismatic planes (Fig.~\ref{FIG: stress/strain small}-(b-3),(c-3),d).

\subsection*{3 MLIPs' up-fitting for nanoscale tensile tests}
MLIP-[0001], MLIP-[10$\overline{1}0$], and MLIP-[$\overline{1}2\overline{1}0$] provide reliable description of TiB$_2$'s response to uniaxial tensile loading at the atomic scale. 
This is indicated by low extrapolation grades ($\gamma\leq5<\gamma_{\text{reliable}}$) as well as accuracy of stress/strain curves, theoretical strength and toughness, and fracture mechanisms consistent with AIMD observations (see previous section).

As a next step, we simulate more realistic tensile tests at the nanoscale, including Poisson’s contraction and continuously increasing the applied strain.
For clarity, we establish the following notation.
\begin{itemize}\setlength\itemsep{0em} 
    \item {\bf{Atomic scale tensile tests}} (presented in the previous section) are uniaxial tensile tests with the following setup:
    \begin{enumerate}[label=(\roman*)]\setlength\itemsep{0em} 
        \item Supercells with $\approx{10^3}$ atoms. Specifically, our supercell has 720 atoms, corresponding to dimensions of $\approx(1.5\times 1.6\times 2.6)\;\mathrm{nm}^{3}$.
        \item A 2\% strain step with a 3~ps equilibration under fixed lattice vectors normal to the loaded direction (i.e. no Poisson's contraction).  
    \end{enumerate}
   
    \item {\bf{Nanoscale tensile tests}} are uniaxial tensile tests with the following setup:
  \begin{enumerate}[label=(\roman*)]\setlength\itemsep{0em}
        \item Supercells with 10$^4$--10$^6$ atoms. Specifically, we use four supercells: $S1$ (12,960 at.), $S2$ (141,120 at.), $S3$ (230,400 at.), and $S4$ (432,000 at.), where $S1\approx{5^3}$~nm$^3$ and $S4\approx{15^3}$~nm$^3$.
        \item Continuous deformation (strain rate 50~\AA/s) and relaxation of lattice vectors normal to the loaded direction (i.e. including Poisson's contraction).  
    \end{enumerate}
\end{itemize}

Employing MLIP-[0001], MLIP-[10$\overline{1}0$], and MLIP-[$\overline{1}2\overline{1}0$] for room-temperature nanoscale tensile tests results in unphysical dynamics (losing atoms) and rapidly increasing extrapolation grades ($\gamma\gg10^3\gg\gamma_{\text{reliable}}$) when approaching the fracture point.
This indicates we have reached length scales where extended defects---absent in atomic scale simulations---dominate materials' deformation.
To enable description of TiB$_2$'s fracture at the nanoscale, our MLIPs require up-fitting (recall Fig.~\ref{FIG: workflow}b).  
Generally, this is a very non-trivial task\cite{hodapp2020operando,freitas2022machine}, since structures causing large $\gamma$ cannot be directly treated by {\it{ab initio}} calculations\footnote{Due to the fact that direct {\it{ab initio}} calculations are unfeasible at the nanoscale, proving that a MLIP predicts the ``correct physics'' is possible only through indirect indications.  
These include the MLIP's uncertainty indication, $\gamma$, and calculations of properties that can be derived also from atomic scale simulations, e.g. lattice parameters or theoretical tensile strength.}.

Below we describe up-fitting steps leading to MLIP-[4] (schematically depicted in Fig.~\ref{FIG: workflow}c), which is a MLIP enabling room-temperature nanoscale tensile tests for TiB$_2$.
\begin{itemize}
    \item We produce {\ul{MLIP-[1]}} by up-fitting MLIP-[0001], where the LS is expanded by final TSs of MLIP-[10$\overline{1}0$] and MLIP-[$\overline{1}2\overline{1}0$]\footnote{Fitting errors (RMSEs) of energies, forces, and stresses do not exceed 3.7~meV/at., 0.12~eV/\AA, and 0.25~GPa, respectively. Extrapolation grades of all {\it{ab initio}} configurations produced so are below $\gamma_{\text{thr}}=2$.
    The final TS has 151 (720-atom) TiB$_2$ configurations, which is less than the individual TSs of MLIP-[0001], MLIP-[10$\overline{1}0$], and MLIP-[$\overline{1}2\overline{1}0$] combined.}. 
    Although MLIP-[1] does not yet suffice to describe TiB$_2$'s fracture during tensile deformation at the nanoscale ($\gamma\gg\gamma_{\text{reliable}}$), it can be applied to simulate atomic scale tensile tests for all the three loading directions.

    In particular, MLIP-[1] yields low extrapolation grades ($\gamma\leq6<\gamma_{\text{reliable}}$) and correctly reproduces fracture mechanisms from AIMD.
    MLIP-[1] also preserves nearly the same accuracy as MLIP-[0001], MLIP-[10$\overline{1}0$], and MLIP-[$\overline{1}2\overline{1}0$]\footnote{Generally, this does not have to be the case. Making a MLIP applicable to a wider spectrum of environments (here, MLIP-[1] can tackle all the three loading conditions) may lead to decreased accuracy for a particular simulation (here, e.g., MLIP-[10$\overline{1}0$] may be in principle more accurate in describing the [10$\overline{1}0$] loading condition).}.
    In terms of (time-averaged) stresses, differences from AIMD values, are 0.13--1.47~GPa, resulting in statistical errors RMSE~$\approx{0.79}$~GPa, R$^{2}$~$\approx{0.9999}$.
    The corresponding stress/strain curves are presented in the Suppl. Mat. (Fig.~S2).
    Additionally, MLIP-[1] provides good accuracy of room-temperature elastic constants, $C_{ij}$ (Tab.~\ref{TAB: EC}), differing from AIMD-calculated values by less than 4.9\%. 
    Although our MLIPs are not targeted to accurate predictions of $C_{ij}$\footnote{A training approach suitable for accurate predictions of $C_{ij}$ has been recently suggested by Ref.~\cite{Method4}.} due to low energy cutoff used in the underlying AIMD training set, the here obtained values further underpin reliability of MLIP-[1]\footnote{
    Furthermore, the only computational report of TiB$_2$'s elastic constants at finite temperatures, besides ours, is Ref.~\cite{xiang2015temperature}, which does not use AIMD but a method based on {\it{ab initio}} calculations of $C_{ij}$ at 0~K and {\it{ab initio}} phonon theory of thermal expansion.}.

\begin{table*}[!htbp]
\centering
\small
\caption{\footnotesize Elastic constants, $C_{ij}$ (in GPa), of $\mathrm{TiB}_{2}$ (at temperature $T$ (K))---calculated using the here-developed MLIP (MLIP-[1])---together with the polycrystalline bulk modulus, $B$ (in GPa), shear modulus, $G$ (in GPa), Young's modulus, $E$ (in GPa), and Poisson's ratio, $\nu$, compared to reference {\it{ab initio}} and experimental (exp.) data. Ref.~\cite{EC1} and Ref.~\cite{EC2} is for TiB$_2$ single and polycrystal, respectively.
AIMD and ML-MD elastic constants were evaluated following Ref.~\cite{sangiovanni2021temperature}, based on a second-order polynomial fit of the [0001], [10$\overline{1}0$], and [$\overline{1}2\overline{1}0$] stress/strain data ($C_{11}$, $C_{12}$, $C_{13}$, $C_{33}$) and of the (0001)$[\overline{1}2\overline{1}0]$, $(10\overline{1}0)[\overline{1}2\overline{1}0]$, and $(10\overline{1}0)[0001]$ shear stress/strain data ($C_{44}$), assuming strains up to 4\%.
For details see the Methodology section.
}
\label{TAB: EC}
\begin{tabular}{ccccccccccccccc}
    \hline
    \hline
     & Nr. of atoms & $T$ & $C_{11}$ & $C_{33}$ & $C_{44}$& $C_{12}$ & $C_{13}$& $E$& $B$ & $G$ & $\nu$ & Source \\
    \hline
    DFT & 3 & 0 &655&461& 260&65&99&582& 253 & 266 &{0.110}&Ref.~\cite{LC4}\\ 
    DFT & 12 & 0 & 654 & 464 & 259 & 76 & 115 & 591 & 263 & 263 &{0.125}&This work\\ 
    AIMD & 720 & 300 & 588& 
    430& 
    252 & 
    79
    &111& 
    547& 
    244 
    & 243 & 
    {0.126 } 
    &This work\\ 
    ML-MD & 720 & 300 & 588 & 409 & 261 & 85 & 98 & 554 & 236 & 246 &{0.113 }&This work\\ 
    Exp. & - & 300 & 660&432&260&48&93&565& 244 & 266 &{ 0.099}&Ref.\cite{EC1}\\ 
    Exp. & - & 300 & 588&503&238&72&84&575& 249 & 255 &{ 0.114}&Ref.~\cite{EC2}\\ 
    \hline
    \hline
\end{tabular}
\end{table*}

    \item We up-fit MLIP-[1] using three different LSs, producing {\ul{MLIP-[2]}}, {\ul{MLIP-[3]}}, and {\ul{MLIP-[4]}}.
    Specifically, MLIP-[2] and MLIP-[3] learn from AIMD snapshots of TiB$_2$  equilibrated at 1200\;K (MLIP-[2]), and sequentially elongated in the [0001] direction until cleavage (MLIP-[3]).
    MLIP-[4] learns from AIMD snapshots of TiB$_2$ elongated by 150\% in the [0001] direction, initializing atoms at ideal lattice sites and equilibrating at 300 and 1200\;K under fixed volume and shape\footnote{Note that all the three LSs are generated at low computational and human time. In particular, we do not ``hand-design'' any material-specific environments or e.g. surfaces, only pull a TiB$_2$ supercell (720-atoms, equilibrated at 300~K and 1200~K) along one of the three lattice vectors.}. Such large strain quickly induces fracture, thus providing additional information for training MLIP on highly deformed lattice environments and surface properties.  

    MLIP-[2], MLIP-[3], and MLIP-[4] all provide reliable description of room-temperature tensile tests at the atomic scale ($\gamma\leq5<\gamma_{\text{reliable}}$).
    The corresponding stress/strain curves nearly overlap with AIMD-calculated ones (Supp. Mat. Fig.~S2) and also fracture mechanisms are correctly reproduced. The accuracy of elastic constants remains the same as for MLIP-[1].
    At the nanoscale, MLIP-[2] and MLIP-[3] exhibit lower $\gamma$ than those obtained by MLIP-[1]. However, the sought improvement ($\gamma\leq\gamma_{\text{reliable}}$) is achieved only by MLIP-[4], which will be used to carry out nanoscale ML-MD simulations of TiB$_2$ deformation.
    
    Besides reliability indication through $\gamma$, MLIP-[4] provides physically sound stress/strain curves and dynamics of nanoscale tensile deformation (illustrated in Supp. Mat. Fig.~S4, and in the following section).
    Additional indication of MLIP-[4]'s reliability are lattice parameters evaluated by equilibrating atomic scale ($\approx{10^3}$ atoms) and nanoscale ($\approx{10^4}$--10$^6$ atoms) TiB$_2$ supercells.
    As shown in Tab.~\ref{TAB: LC}, the values do not differ by more than 0.01\%.
\end{itemize}

\begin{table}[!htbp]
\centering
\footnotesize
\caption{\footnotesize Lattice constants, $a$ and $c$ (in \AA), of $\mathrm{TiB}_{2}$ at temperature $T=\{0,300,1200\}$\;K calculated using the here-developed ML potential (MLIP-[4]), compared to reference {\it{ab initio}} and experimental (exp.) values (Ref.~\cite{LC2} and Refs.~\cite{LC1,LC3} are for TiB$_2$ powder and thin films, respectively).}
\label{TAB: LC}
\begin{tabular}{ccccccc}
    \hline
    \hline
     & Nr. of atoms & $T$ & $a$ & $c$ & Source \\
    \hline
    DFT & 3  & 0 & 3.038 & 3.234 & Ref.~\cite{LC4}\\
    DFT & 12  & 0 & 3.033 & 3.227 & This work\\
    AIMD & 720 & 300 &3.035&3.218& This work\\
    ML-MD & 720 & 300 &3.036&3.218& This work\\
    ML-MD & (13--430)$\cdot10^3$ & 300 & 3.036 & 3.217 & This work\\     
    Exp. & - & 300 & 3.032&3.229& Ref.~\cite{LC2}\\
    Exp. & - & 300 & 3.029&3.229& Ref.~\cite{LC1}\\
    Exp. & - & 300 & 3.021&3.230& Ref.~\cite{LC3}\\
    AIMD & 720 & 1200 & 3.056& 3.249 & This work\\
    ML-MD & 720 & 1200 & 3.056& 3.249 & This work\\
    ML-MD & (13--430)$\cdot10^3$ & 1200 & 3.047 & 3.239 & This work\\
\hline
\hline
\end{tabular}
\end{table}

To offer some understanding of why MLIP-[4] enables nanoscale tensile tests, one needs to analyse the corresponding training set, TS(MLIP-[4]).
In Fig.\ref{FIG: training sets}, we visualize  selected characteristics of TS(MLIP-[4]) in comparison to the training set of MLIP-[1], TS(MLIP-[1]), where the latter is not applicable to simulate TiB$_2$'s fracture at the nanoscale.
The radial distribution function (RDF, Fig.\ref{FIG: training sets}a) and bond angle distribution analysis (Suppl. Mat., Fig.S6) suggest minor geometrical differences between structures contained in TS(MLIP-[1]) and TS(MLIP-[4]).
Their total energy and stress distribution, however, differ significantly (Fig.\ref{FIG: training sets}b).
In particular, TS(MLIP-[4]) contains atomic configurations with higher total energy and higher total energy in combination with higher stress in principal crystallographic axes, which are missing in TS(MLIP-[1]).
An illustration of structures from the two training sets is given in Fig.\ref{FIG: training sets}c.
The chosen snapshots indicate that TS(MLIP-[4]) provides a variety of atomic environments relevant for simulations of non-stoichiometry, locally amorphous regions, and surfaces, which are likely to be helpful also for simulations of extended defects nucleating due to high strains during tensile tests. 

\begin{figure}[!htbp]
\centering
\includegraphics[width=0.45\textwidth]{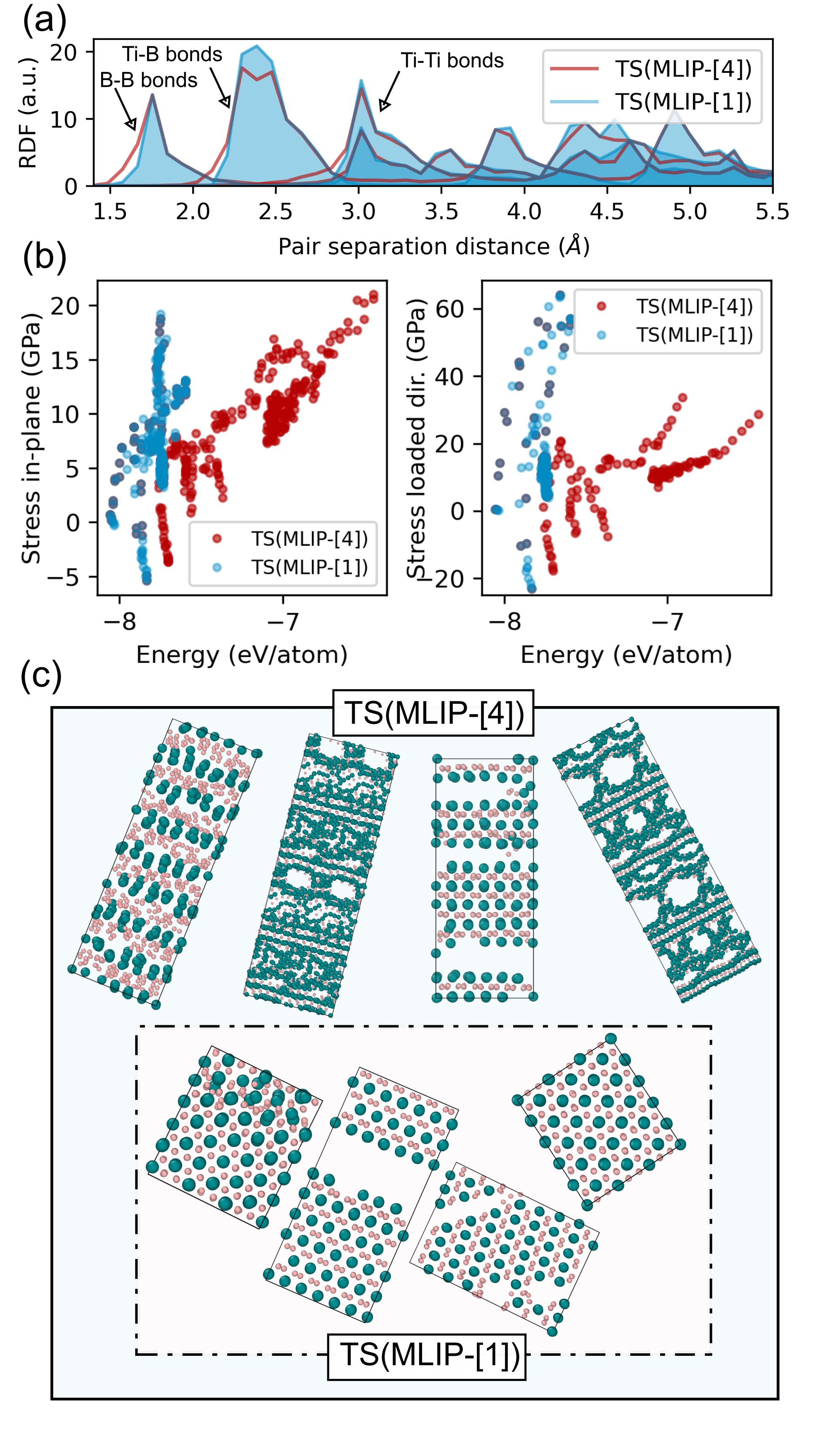}
\caption{\footnotesize 
Qualitative differences between training sets (TSs) of the here-developed ML potentials (MLIP-[1], MLIP-[4]). Only MLIP-[4] is suitable for nanoscale ML-MD tensile tests.
(a) Radial distribution function (RDF, with 5.5~\AA\ cutoff) for B--B, Ti--B, and Ti--Ti bonds (integrated over all configurations).
(b) Stress components (in-plane and in the loaded direction) vs. total energy of all configurations in the training set.
(c) Snapshots of representative structures from the two training sets.
}
\label{FIG: training sets}
\end{figure}

\subsection*{4 Size effects in tensile response of TiB$_2$}
Equipped with the above developed MLIP-[4], in this section we discuss TiB$_2$'s response to room-temperature uniaxial tensile loading from atomic to nanoscale.
Recall that an important difference between our atomic and nanoscale tensile tests is that the former disregard Poisson's contraction.
The stress/strain curves calculated at room temperature are depicted in Fig.~\ref{FIG: stress/strain large}.
For strains below $\approx{10}$\%, stresses for the respective loading direction ([0001], $[10\overline{1}0]$, and $[\overline{1}2\overline{1}0]$) almost perfectly overlap regardless the supercell size ($\approx{10^3}$--10$^6$ atoms).
Such overlap indicates consistency in elastic constants derived from atomic and nanoscale models\footnote{Note that in order to obtain the $C_{44}$ elastic constant, we would need to perform additional shear simulations, which we did only at the atomic scale (recall Tab.~\ref{TAB: EC}).}.
Furthermore, also elastic isotropy of the basal plane is size independent. The isotropy is suggested by nearly the same initial slope of stress/strain curves for $[\overline{1}2\overline{1}0]$ and $[10\overline{1}0]$ elongation and in line with experimental reports for hexagonal crystals~\cite{paul2021plastic}.

Due to Poisson contraction, however, differences in stress/strain curves emerge beyond the linear-elastic regime.
A shrinkage 
of the lattice parameters normal to the applied tensile strain yields Poisson's ratio ($\nu\approx{0.127}$ \footnote{ The Poisson contraction was calculated as $\nu=-\frac{d \varepsilon_{\text {compressed }}}{d \varepsilon_{\text {elongated }}}$, where the $d \varepsilon_{\text {compressed}}$ ($d \varepsilon_{\text {elongated }}$) is the lattice parameter shrinkage (increment) orthogonal (parallel) to the loading direction. The presented value is an average of Poisson's ratios for both in-plane directions.}) consistent with the value obtained from elastic constants($\nu\approx{0.113}$ , see Tab.~\ref{TAB: EC}), thus underpinning reliability of MLIP-[4].
Approaching the fracture point, differenc es between TiB$_2$'s tensile behavior at atomic and nanoscale become more apparent\footnote{Even more significant size effects may be expected for materials with strain-activated plasticity mechanisms.}.  
These can be illustrated by ideal strengths and toughness moduli (Tab.~\ref{TAB: mechanical properties}). Although theoretical strength and toughness exhibit qualitative differences in their directional dependence at the atomic scale and nanoscale, these properties are well saturated for all nanoscale supercells (S1--S4).
Specifically, the $[\overline{1}2\overline{1}0]$ direction exhibits the highest tensile strength ($\approx{56}$~GPa), followed by the [0001] direction ($\approx{54}$~GPa), and the $[10\overline{1}0]$ direction ($\approx{51}$~GPa)\footnote{At the atomic scale, in contrast, TiB$_2$'s tensile strength is the highest in the $[\overline{1}2\overline{1}0]$ direction ($\approx{64}$~GPa), followed by $[10\overline{1}0]$ ($\approx{55}$~GPa) and [0001] ($\approx{53}$~GPa).}.
For comparison, the hardness reported for TiB$_2$ thin films, typically [0001]-textured, is 41--53~GPa\cite{H1,H2,H3}.
The [0001] direction exhibits the highest toughness modulus ($\approx{4.80}$~GPa), followed by the $[\overline{1}2\overline{1}0]$ direction ($\approx{3.37}$~GPa), and the $[10\overline{1}0]$ direction ($\approx{2.78}$~GPa)\footnote{Atomic-scale ML-MD indicates nearly the same [0001] and $[\overline{1}2\overline{1}0]$ tensile toughness ($\approx{4.32}$~GPa) which is above that reached during the $[10\overline{1}0]$ tensile test ($\approx{3.11}$~GPa).}.

\begin{figure*}[!htbp]
\centering
\includegraphics[width=1\textwidth]{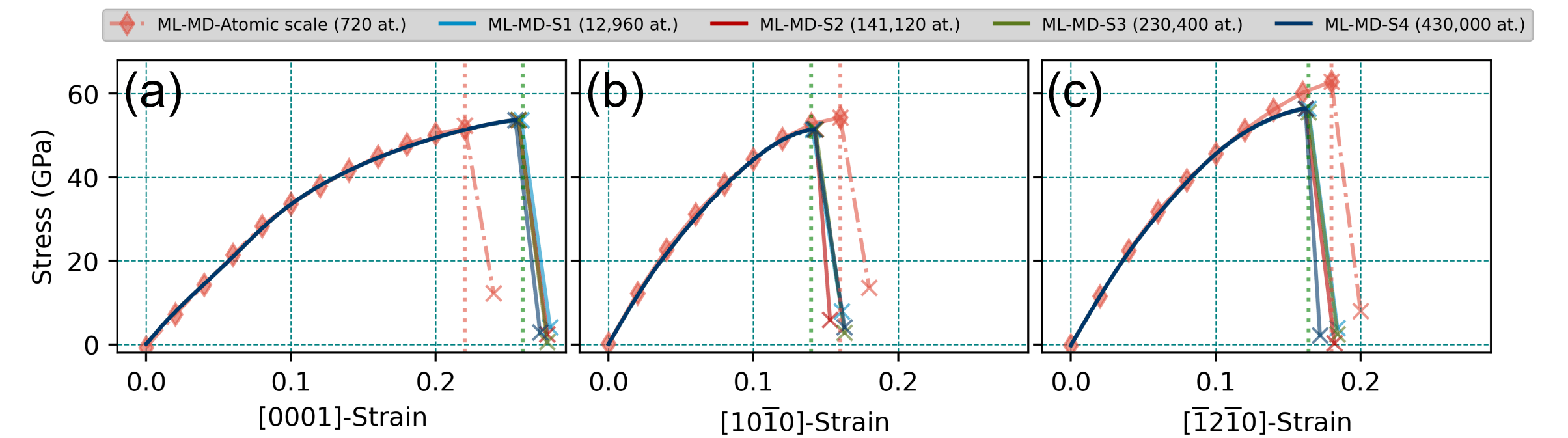}
\caption{\footnotesize
ML-MD stress/strain curves (obtained by MLIP-[4]) for TiB$_2$ subject to (a) $[0001]$, (b) $[10\overline{1}0]$, and (c) $[\overline{1}2\overline{1}0]$ tensile deformation at 300~K.
Only the stress component in the loaded direction is plotted.
The orange diamonds correspond to atomic scale ML-MD simulations, while the solid lines correspond to nanoscale ML-MD simulations, as defined at the beginning of this section.}
\label{FIG: stress/strain large}
\end{figure*}

\begin{table*}[!htbp]
\centering
\small
\caption{\footnotesize Mechanical properties of various TiB$_2$ supercells derived from room-temperature ML-MD stress/strain data (Fig.~\ref{FIG: stress/strain large}) together with 1200\;K results for comparison.}
\label{TAB: mechanical properties}
\resizebox{\textwidth}{!}{
\begin{tabular}{ccccccccccccccc}
    \hline
    \hline
    Nr. of atoms & T (K)& \multicolumn{3}{c}{Dimensions (nm) }& \multicolumn{3}{c}{Strength (GPa)} & \multicolumn{3}{c}{Toughness (GPa)}&\multicolumn{3}{c}{ Fracture strain (\%)}\\
      { } &{ } & $a$ & $b$ & $c$ & $[0001]$ & $[10\overline{1}0]$ & $[\overline{1}2\overline{1}0]$  & $[0001]$ & $[10\overline{1}0]$ & $[\overline{1}2\overline{1}0]$ & $[0001]$ & $[10\overline{1}0]$ & $[\overline{1}2\overline{1}0]$\\
    \hline
    720 & 300  & 1.51 & 1.58 & 2.57 &52.72&55.01&63.69&4.33&3.11&4.32&22.0 &16.0 &18.0\\
    12,960 ($S1$) &300  & 4.55 & 4.73 & 5.15 & 53.71&51.36&56.40&4.83&2.77&3.38&26.4 &14.3 &16.8\\
    141,120 ($S2$) &300  & 10.63 & 11.05 & 10.30 & 53.69&51.43&56.41&4.81&2.78&3.37&26.4 &14.3 &16.8\\
     230,400 ($S3$) &300  & 12.14 & 12.63 & 12.87 & 53.71&51.43&56.39&4.82&2.78&3.37&26.3 &14.3 &16.8\\
     432,000 ($S4$)  &300  & 15.18 & 15.79 & 15.45 &53.64&51.47&56.42&4.80&2.78&3.37&26.2 &14.2 &16.8\\
    720 & 1200  & 1.53 & 1.59 & 2.60 &43.87&45.21&51.35&3.27&2.29&3.07&20.0 &16.0 &16.0\\
    432,000 ($S4$) & 1200 & 15.26 & {15.86}& {15.52}& {43.27}& {41.31}& {44.72}& {3.45}& {1.96}& {2.32}& {21.6}& {12.8}& {14.1}\\
\hline
\hline
\end{tabular}}
\end{table*}

Besides characterizing directional dependence of tensile strength and toughness in dislocation-free monocrystals, nanoscale simulations also provide insights into crack nucleation and growth mechanisms.  
This can be illustrated---and contrasted with atomic scale ML-MD---using the example of [0001] tensile test (Fig.~\ref{FIG: snapshots from [001] tensile}).
At the atomic scale, all atoms in TiB$_2$ vibrate close to their ideal lattice sites until a sudden brittle cleavage induces formation of two surfaces almost perfectly parallel with (0001) basal planes (Fig.~\ref{FIG: snapshots from [001] tensile}, row 1).
At the nanoscale, fracture is initiated by opening of voids accompanied by local necking which produces lattice re-orientations (Fig.~\ref{FIG: snapshots from [001] tensile}, row 2 and 3). 
Rapid void coalescence and fraying of ligaments results in corrugated fractured surfaces, predominantly with (0001) orientation.
Following the stress release, inner parts of the crystal relax back to the ideal TiB$_2$ lattice sites.

The $S1$ supercell yields in only one region (Fig.~\ref{FIG: snapshots from [001] tensile}, row 2).
The larger $S2$, $S3$, and $S4$ supercells---where only $S4$ is depicted in Fig.~\ref{FIG: snapshots from [001] tensile}---do not fracture in two pieces but show few-nm-size cracks inside.
Here, fracture surfaces do not align only with the basal planes but also with the \{10$\overline{1}$1\} first order pyramidal planes (recall the notation in Fig.~\ref{FIG: stress/strain small}d).
Volumetric strain analysis (Fig.~\ref{FIG: snapshots from [001] tensile}d,e) highlights locally increased tensile strain concentration surrounding small voids (see TiB$_2$ slice at $\approx{27}$\% strain) due to decreased interplanar spacings between Ti and B layers (predominantly due to [0001] compression) above and below the voids.
The largest $S4$ supercell allows for crack propagation under different directions, thus, provides a certain statistical viewpoint missing in the $S1$ supercell and at the atomic scale.

\begin{figure*}[!htbp]
\centering
\includegraphics[width=1.00\textwidth]{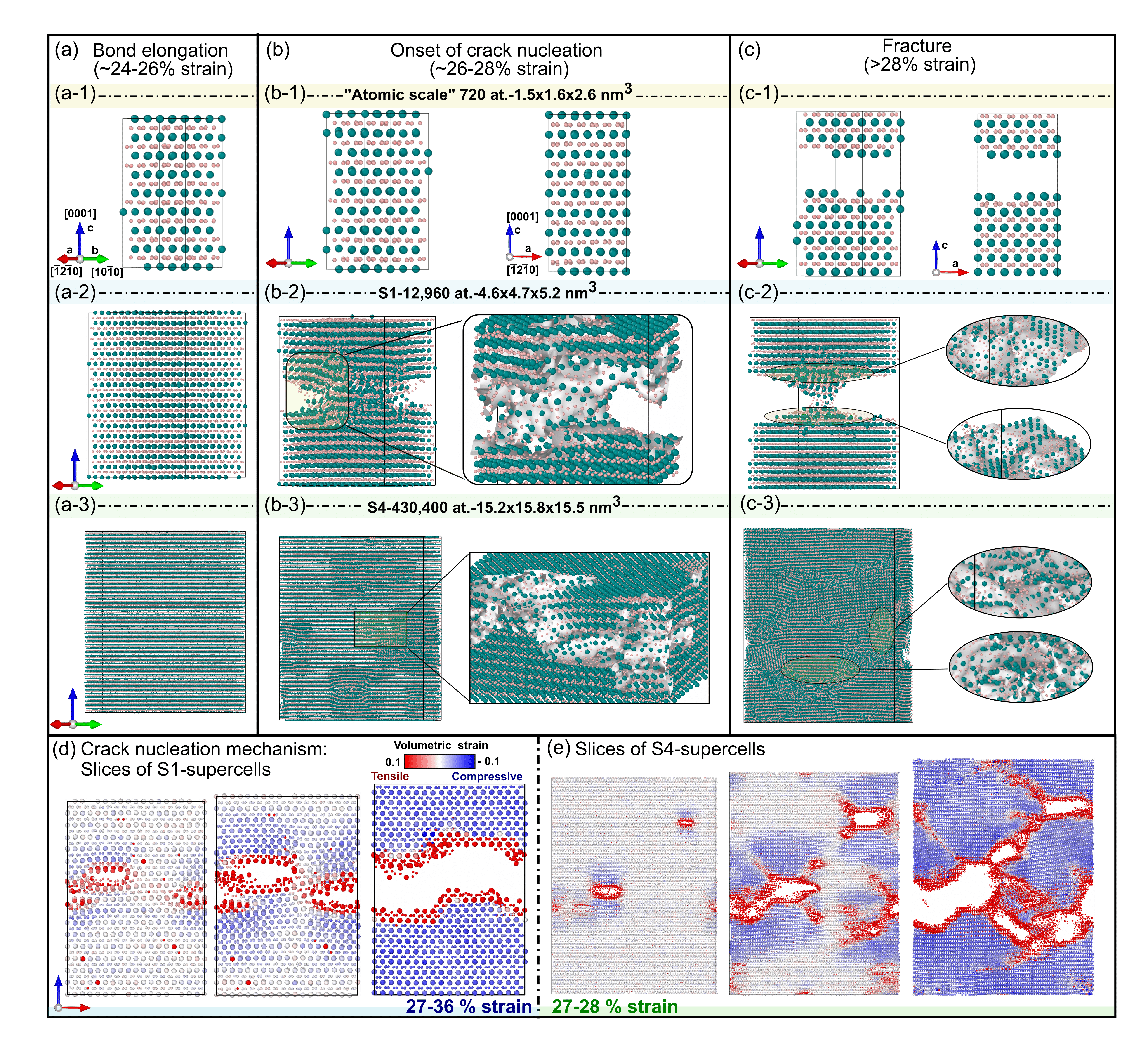}
\caption{
\footnotesize
Illustration of size effects in (room-temperature) ML-MD [0001] tensile tests for TiB$_2$ at key deformation stages: (a) {\bf{bond elongation}} in the loaded direction, (b) {\bf{onset of crack nucleation}}, and (c) {\bf{fracture}}.
Thin slices of the nanoscale (d) $S1$ and (e) $S4$ supercells color-coded based on volumetric strain (using the corresponding equilibrium structure as reference).
Red (blue) regions denote high tensile (compressive) strain.
}
\label{FIG: snapshots from [001] tensile}
\end{figure*}

For the $[10\overline{1}0]$ tensile test, size effects in fracture mechanisms are compared in Fig.~\ref{FIG: snapshots from [010] tensile}.
At the atomic scale, two voids open diagonally across Ti/B layers (Fig.~\ref{FIG: snapshots from [010] tensile}, row 1).
At the nanoscale, we observe nucleation of V-shaped cracks, as illustrated for the $S1$ and the $S4$ supercell (Fig.~\ref{FIG: snapshots from [010] tensile}, row 2 and 3, and panels c, d), where $S4$ additionally reveals lattice rotation near the V-shaped defects.
We infer that loading in the direction of strong covalent B--B bonds most often induces crack deflection and fracture at \{11$\overline{2}2\}$ family of surfaces parallel to the second order pyramidal planes.

\begin{figure*}[!htbp]
\centering
\includegraphics[width=1.0\textwidth]{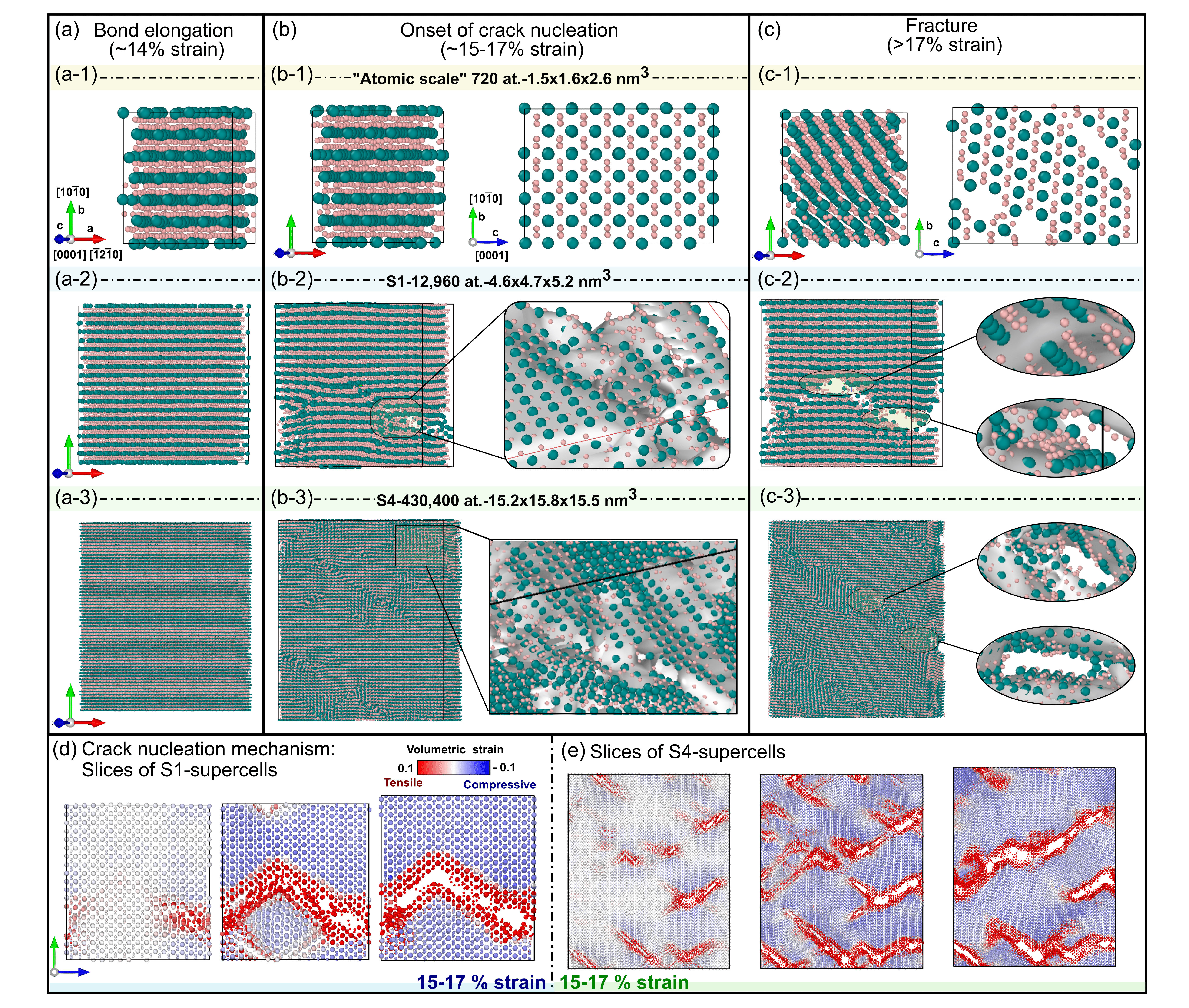}
\caption{\footnotesize
Illustration of size effects in (room-temperature) ML-MD $[10\overline{1}0]$ tensile tests for TiB$_2$ at key deformation stages: (a) {\bf{bond elongation}} in the loaded direction, (b) {\bf{onset of crack nucleation}}, and (c) {\bf{fracture}}.
Thin slices of the nanoscale (d) $S1$ and (e) $S4$ supercells color-coded based on volumetric strain (using the corresponding equilibrium structure as reference).
Red (blue) regions denote high tensile (compressive) strain.
}
\label{FIG: snapshots from [010] tensile}
\end{figure*}

Changing to the $[\overline{1}2\overline{1}0]$ tensile deformation, atomic scale simulations predict fracture along \{10$\overline{1}$0\} prismatic planes (Fig.~\ref{FIG: snapshots from [100] tensile}, row 1).
This is underpinned also by nanoscale ML-MD (Fig.~\ref{FIG: snapshots from [100] tensile}, row 2 and 3), suggesting that crack growth is most often orthogonally and diagonally across Ti/B layers (see the dashed line with arrow in Fig.~\ref{FIG: snapshots from [100] tensile}e).

\begin{figure*}[!htbp]
\centering
\includegraphics[width=1.0\textwidth]{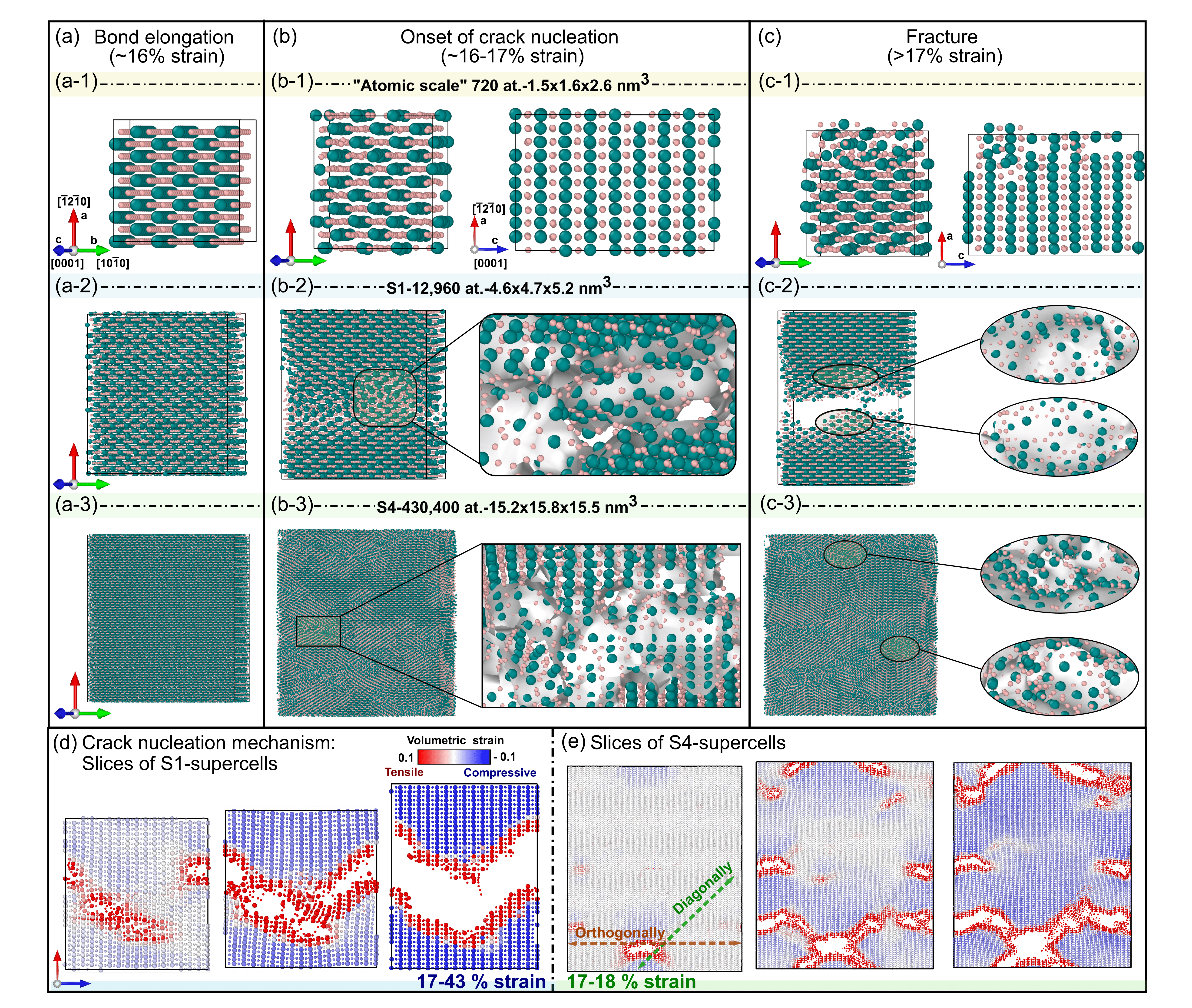}
\caption{\footnotesize
Illustration of size effects in (room-temperature) ML-MD $[\overline{1}2\overline{1}0]$ tensile tests for TiB$_2$ at key deformation stages: (a) {\bf{bond elongation}} in the loaded direction, (b) {\bf{onset of crack nucleation}}, and (c) {\bf{fracture}}.
Thin slices of the nanoscale (d) $S1$ and (e) $S4$ supercells color-coded based on volumetric strain (using the corresponding equilibrium structure as reference).
Red (blue) regions denote high tensile (compressive) strain.
}
\label{FIG: snapshots from [100] tensile}
\end{figure*}

\subsection*{5 Other loading conditions and MLIP's transferability}
Targeted applications of our MLIP (MLIP-[4]) are atomic to nanoscale simulations of TiB$_2$ subject to room-temperature uniaxial tensile loading.
Although the training set only contained snapshots of a 720-atom TiB$_2$ supercell, with one of the three lattice vectors elongated with respect to the equilibrium length at 300~K, Fig.~\ref{FIG: training sets}c indicates variety of atomic environments relevant for simulations of non-stoichiometry, locally amorphous regions, or e.g. surfaces.

Here we discuss transferability of our MLIP to simulations for which it has not been explicitly trained.
Accuracy of the predicted observables (e.g. shear strengths or surface energies) is presented in the context of extrapolations grades, allowing to identify types of configurations beneficial for further up-fitting, thus, broadening the MLIP's  applicability.  

\begin{itemize}
\item {\bf{High-temperature tensile deformation of TiB$_2$.}}
    Since TiB$_2$ is a UHTCs (see the Introduction), modelling its mechanical behaviour at extreme temperatures is of high practical relevance.
    Here we choose 1200~K which is close to the highest anti-oxidation temperature of Ti reported experimentally~\cite{jiao2018progress}. 
    
    Atomic scale [0001], $[10\overline{1}0]$, and [$\overline{1}2\overline{1}0$] tensile tests at 1200\;K show excellent quantitative agreement with AIMD simulations at the same temperature (see Suppl. Mat., Fig.~S5).
    Specifically, differences from AIMD-calculated stresses are 0.01--3.54~GPa, resulting in statistical errors RMSE~$\approx{1.85}$~GPa, R$^{2}$~$\approx{0.9997}$.
    Extrapolations grades indicate reliable extrapolation ($\gamma\leq7<\gamma_{\text{reliable}}$).  

    TiB$_2$'s theoretical tensile strength at 1200\;K decreases by about 17--19\% compared to 300\;K.
    For tensile toughness, our simulations predict $\approx{25}$\% decrease. 
    Fracture mechanisms remain qualitatively unchanged with respect to 300~K.

\item {\bf{Room-temperature shear deformation of TiB$_2$}}. 
    Simulations of shear deformation provide useful insights for understanding of how dislocations nucleate and move in generally brittle UHTCs\cite{dub2017mechanical,lei2019synthesis}. 
    Furthermore, as diborides typically crystallize in layered structures ($\alpha$, $\gamma$, $\omega$) which can be viewed as different stackings of the transition metal planes, shearing may also induce transformation toughening\cite{leiner2023energetics}.
    Following Ref.~\cite{zhang2010ideal}, we choose to simulate the (0001)$[\overline{1}2\overline{1}0]$, $(10\overline{1}0)[\overline{1}2\overline{1}0]$, and $(10\overline{1}0)[0001]$ shear deformation\footnote{There are various slip systems experimentally observed in diborides depending on temperature. For an overview, please see Ref.~\cite{paul2021plastic}.}.
        
    Stress evolution during atomic scale shear deformation (Fig.~\ref{BMCS}a) agrees well with equivalent AIMD simulations.
    This is particularly the case for strains below $\approx{20}\%$, where stresses differ from AIMD values by 0.01--5.08~GPa (yielding statistical errors RMSE~$\approx{3.72}$~GPa, R$^{2}$~$\approx{0.9993}$) and $\gamma$ is close to reliable extrapolation ($\gamma<14$). 
    Considering that the training set did not contain any sheared supercells, this is a plausible result.
    Shear strains above $\approx{20}\%$ induce notably larger discrepancies in stresses (differing from AIMD by 5.36--8.51~GPa) and increased $\gamma$ ($\gamma\leq 26$). 
    The main reason is that lattice slip---responsible for (partial) stress release subject to shearing---does not occur exactly at the same strain step, although the mechanism is identical (Fig.~\ref{BMCS}b--d)\footnote{Note that if repeated many times, AIMD shear simulations would not exhibit lattice slip exactly at the same strain step either.}.
   
    The predicted shear strengths (49, 57, and 51\;GPa for the (0001)$[\overline{1}2\overline{1}0]$, $(10\overline{1}0)[\overline{1}2\overline{1}0]$, and $(10\overline{1}0)[0001]$ deformation, respectively) are $\approx{8}\%$ below AIMD values (58, 72, and 68\;GPa).      
    Shear-induced structural changes correctly reproduce AIMD observations.
    Specifically, a basal plane slip is activated subject to $\approx{24}$\% (0001)[$\overline{1}2\overline{1}0$] shear strain, restoring nearly ideal TiB$_2$ lattice sites (Fig.~\ref{BMCS}b).
    The $(10\overline{1}0)[\overline{1}2\overline{1}0]$ shear (Fig.~\ref{BMCS}c) induces plastic deformation, first via $(\overline{1}2\overline{1}0)[10\overline{1}0]$ slip, which occurs at $\approx{30}$\%, and then via $(10\overline{1}0)[\overline{1}2\overline{1}0]$ slip at $\approx{50}$\% strain. Both mechanisms are accompanied by displacements of Ti and B atoms from the ideal TiB$_2$ lattice sites\footnote{ 
    The phenomenon may be rationalized by high energetic costs of slip in the (tilted) $[\overline{1}2\overline{1}0]$ direction, which would require breaking strong covalent B--B bonds.}.
    The $(10\overline{1}0)[0001]$ shear (Fig.~\ref{BMCS}d) causes slipping along the (tilted) $[10\overline{1}0]$ direction ($\approx{26}$\% strain) and the $[0001]$ direction ($\approx{50}$\% strain), where the latter results in significant displacements of B atoms.

     Overall, we infer that MLIP up-fitting would benefit from severely sheared configurations.
     Already the current potential (MLIP-[4]), nonetheless, suffices to estimate shear strengths and stress release mechanisms at the atomic scale. 
\end{itemize}

\begin{figure}[!htbp]
\centering
\includegraphics[width=0.5\textwidth]{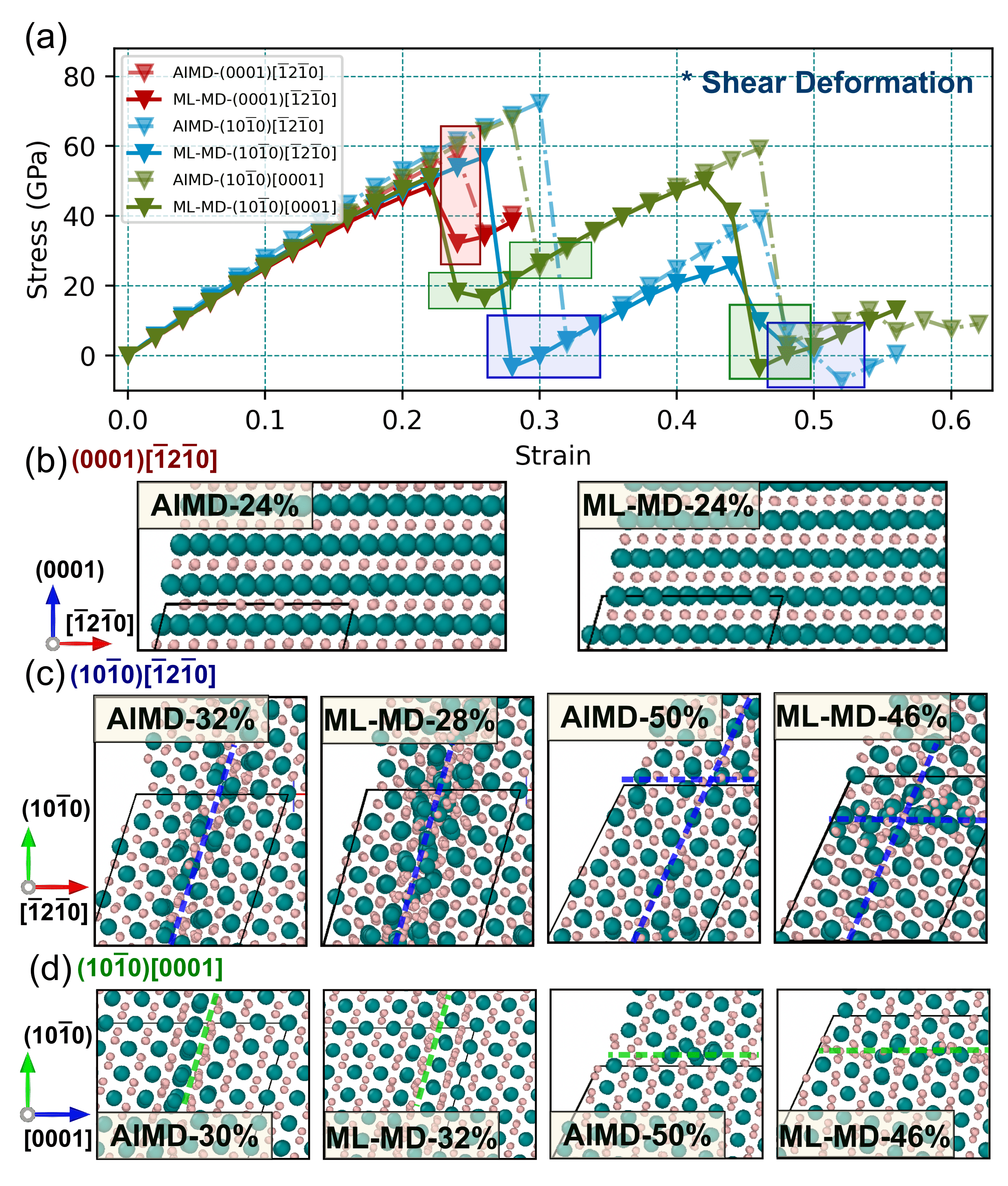}
\caption{\footnotesize Transferability of the here-developed MLIP (MLIP-[4]) to atomic scale simulations of room-temperature shear deformation. 
(a) Comparison of AIMD and ML-MD stress/strain curves for TiB$_2$ subject to (0001)$[\overline{1}2\overline{1}0]$, $(10\overline{1}0)[\overline{1}2\overline{1}0]$, and $(10\overline{1}0)[0001]$ shear.
(b, c, d) Selected snapshots at strain steps marked by shaded rectangles in (a). The dashed lines guide the eye for the slip direction described in the text.
}
\label{BMCS}
\end{figure}

\begin{itemize}
\item {\bf{Room-temperature volumetric compression of TiB$_2$}}.
    Training on snapshots of compressed structures may be important not only in obvious cases, such as e.g. simulations of uniaxial compression or nanoindentation, but also in order to correctly account for Poisson's contraction during tensile deformation.
    For TiB$_2$, the relatively low Poisson's ratio (Tab.~\ref{TAB: EC}) induces rather small compression, accurately reproduced by MLIP-[4] at the nanoscale\footnote{Specifically, Poisson's ratio derived from lattice parameter shrinkage during nanoscale tensile tests quantitatively agrees with that calculated from AIMD elastic constants (Tab.~\ref{TAB: EC}).}.
    Here we simulate severe (room-temperature) volumetric compression---shrinking {\it{all}} lattice vectors by up to 12\%---to illustrate rapidly growing $\gamma$ and how this can be improved by up-fitting.
       
    As shown in Fig.~\ref{FIG: compression}, compression-induced stresses along main crystallographic directions are indeed {\it{extreme}}.
    In AIMD, they exceed 50\;GPa for a 5\% compression and reach $\approx{150}$\;GPa for a 10\% compression.    
    Our MLIP yields satisfactory agreement with AIMD for volumetric compression of 1--2\%, with stress tensor components differing from {\it{ab initio}} values by less than 1.83\;GPa (9.57\%) and $\gamma$ indicating reliable extrapolation ($\gamma\leq10=\gamma_{\text{reliable}}$).
    Further compression (of each lattice vector) up to 10\%, however, causes increasing deviations from AIMD stresses and$\gamma\approx{10^2}$--10$^3$.  
    
    We illustrate the effect of up-fitting by producing a new MLIP (MLIP-[4]$_{Plus}$) which learns from AIMD snapshots of a 10\% volumetrically-compressed TiB$_2$ (added to the LS of MLIP-[4]).
    This not only greatly improves accuracy for the 10\% compression---stress differences are maximum 0.79\;GPa (0.48\%) and $\gamma\leq2$---but also within the entire compression range (see red data points in Fig.~\ref{FIG: compression}).
\end{itemize}
    
\begin{figure}[!htbp]
    \centering
    \includegraphics[width=0.49\textwidth]{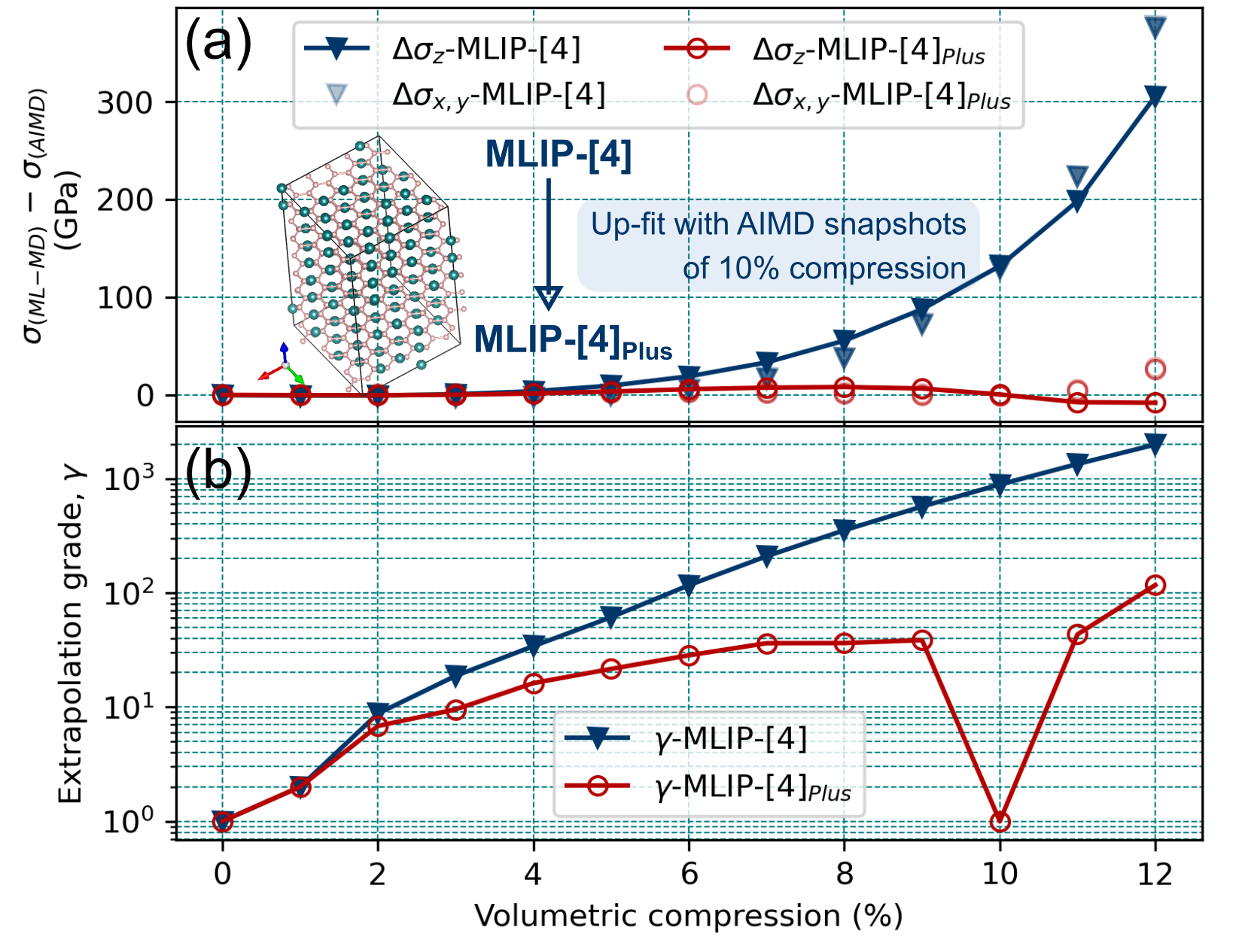}
    \caption{\footnotesize Transferability of the here-developed MLIP (MLIP-[4]) to atomic scale room-temperature simulations of volumetric compression.
   (a)~Differences in main stress tensor components ($\sigma_{x,y}$, in the basal plane, and $\sigma_z$ in the [0001] direction) between ML-MD and AIMD simulations, plotted as a function of the compression percentage.
   (b)~Evolution of extrapolation grades. 
   Blue and red data point correspond to ML-MD simulation with different MLIPs, MLIP-[4] and MLIP-[4]$_{Plus}$.
   }
\label{FIG: compression}
\end{figure}

\begin{itemize}
\item {\bf{Surface energies of TiB$_2$}}.
    Surfaces are undoubtedly relevant for simulations of fracture.
    For routine MLIP development, however, it is desirable to minimize human time handcrafting material-specific surfaces.
    The here-suggested training data generation is easily automated and may already include atomic environments relevant for calculations of low-energy surfaces (recall Fig.~\ref{FIG: stress/strain small}c).
    
    Methodologically, calculations of finite-temperature surface energies, $E_{\text{surf}}$, are a non-trivial task.
    We stick to a simple 0\;K calculation of stoichiometric (1:2 Ti-to-B) surfaces via molecular statics (MS).
    The $E_{\text{surf}}$ values predicted by ML-MD differ from {\it{ab initio}} calculations by 0.03\;J/m$^2$ (1.40\%) for $E_{\text{surf}}$(0001), 0.04\;J/m$^2$ (1.68\%) for $E_{\text{surf}}$$(\overline{1}2\overline{1}0)$), and 0.13\;J/m$^2$ (5.75\%) for $E_{\text{surf}}$$(10\overline{1}0)$.
    This is a very good agreement, underlined also by low extrapolation grades ($\gamma\leq12$)\footnote{
    For accurate surface energy calculations, however, relevant surface structures should be explicitly trained on using accurate-enough {\it{ab initio}} dataset (in contrast to our philosophy of large but less accurate training configurations).}. 
\end{itemize}

\begin{itemize}
\item {\bf{Off-stoichiometric TiB$_2$ structures and other phases.}}
    Our MLIP was trained to snapshots of TiB$_2$ (AlB$_2$-type phase, P6/mmm) with a perfect stoichiometry (speaking of the entire supercell).
    Visualization of the training set (Fig.~\ref{FIG: training sets}c), however, indicates presence of atomic environments with various Ti-to-B ratios as well as bond lengths and angles different from those in TiB$_2$. 
    This may be useful for simulations of e.g. vacancy-containing TiB$_2$ structures commonly reported by synthesis~\cite{magnuson2022review} or other phases in the Ti--B phase diagram, although it is quite a stretch from the intended applicability of our MLIP.

    To investigate transferability to other phases, we use MS to find the ground-state of known phases from the Ti--B phase diagram\cite{MaP}: Ti$_2$B (tetragonal, I4/mcm), Ti$_3$B$_4$ (orthorhombic, Immm), and TiB (orthorhombic Pnma) at 0~K\footnote{The supercell sizes are always $\approx{700}$ atoms, i.e. comparable to that used for TiB$_2$.}. 
    Additionally, we equilibrate the TiB$_2$ phase with Ti, B, or combined Ti and B vacancies: Ti$_{36}$B$_{71}$, Ti$_{35}$B$_{72}$, and Ti$_{35}$B$_{70}$.
    For all calculations, extrapolation grades ($\gamma\approx{10^2}\text{--}10^4$) are far beyond {\it{reliable}} extrapolation.
    In terms of total energies ($E_{\text{tot}}$), and lattice parameters ($a$, $b$, $c$), the largest deviation from {\it{ab inito}} values is exhibited by Ti$_3$B$_4$ (yielding 10\% and 2.5\% deviation from {\it{ab inito}} $E_{\text{tot}}$ and $c$, respectively).

\begin{table}[!htbp]
\centering
\small
\caption{\footnotesize Zero Kelvin surface energies, $E_{\text{surf}}$ (in $\mathrm{J}/\mathrm{m}^2$), of low-index $\mathrm{TiB}_{2}$ surfaces predicted with the here-developed MLIP (MLIP-[4]), compared to reference {\it{ab initio}} values.}
\label{TAB: surface energies}
\begin{tabular}{ccccccc}
    \hline
    \hline
    & Surface & $E_{\text{surf}}$ & Source \\
    \hline
    ML-MS & (0001) & 3.80 & This work\\
    DFT & (0001) & 3.80 & This work\\
    DFT & (0001) & 4.50--4.72  & Ref.~\cite{SE-1}\\
    DFT & (0001) & 4.22--4.24 & Ref.~\cite{sun2019general}\\
    DFT & (0001) & 4.14  & Ref.~\cite{SE-2}\\
    DFT & (0001) & $\approx{4.20}$ & Ref.~\cite{SE-3}\\
    \hline
    ML-MS & (10$\overline{1}$0) & 3.98 & This work\\ 
    DFT & (10$\overline{1}$0) & 4.12 & This work\\
    DFT & (10$\overline{1}$0) & 4.20--4.83 & Ref.~\cite{sun2019general}\\ 
    DFT & (10$\overline{1}$0) & $\approx{4.10}$ & Ref.~\cite{SE-3}\\
    \hline
    ML-MS & ($\overline{1}2\overline{1}0$) & 3.42  & This work\\
    DFT & ($\overline{1}2\overline{1}0$) & 3.57  & This work\\
    DFT & ($\overline{1}2\overline{1}0$) & 5.06  & Ref.~\cite{sun2019general}\\
    DFT & ($\overline{1}2\overline{1}0$) & $\approx{4.02}$ & Ref.~\cite{SE-3}\\
\hline
\hline
\end{tabular}
\end{table}
    
    Simulations of other stoichiometries and phases therefore require up-fitting (not necessarily due to poor accuracy but especially due to high uncertainty, $\gamma\gg\gamma_{\text{reliable}}$).
    To illustrate the up-fitting effect, MLIP-[4] learns from additional {\it{ab initio}} snapshots: from 0~K equilibration of a 780-atom Ti$_3$B$_4$ supercell.
    Prior to up-fitting, equilibration of Ti$_3$B$_4$ yields $\gamma\geq10^4$.
    Afterwards, $\gamma\leq 5<\gamma_{\text{reliable}}$ and  $E_{\text{tot}}$, $a$ and $c$ deviate from {\it{ab inito}} values by 4.18\%, 0.07\%, and 0.73\%\footnote{We can also compare the structures relaxed by {\it{ab initio}} calculations and MS, by calculating dfferences between the corresponding atomic positions. Prior to up-fitting, square root of the maximum difference (of each corresponding atom position in Ti$_{3}$B$_{4}$, 780 at.) is 6.77~\AA, which decreases to 6.60 after up-fitting. 
    For comparison, initial TiB$_{2}$ (720 at.) configuration would have maximum difference of 2.28~\AA.
    }
    
\end{itemize}


\section*{Conclusions}
We proposed a strategy for the development of MLIPs suitable to portray intrinsic tensile strength, toughness, and fracture mechanisms of monocrystals from atomic scale ($\approx{10^3}$ atoms, accessible by {\it{ab initio}} methods) to nanoscale ($\approx{10}^4$--10$^6$ atoms). 
Training data generation, fitting, and validation procedure were illustrated using the moment tensor potential (MTP) formalism.
Material-wise, TiB$_2$ ceramic served as a model system.
The MLIP was used to simulate room-temperature uniaxial tensile deformation of TiB$_2$ at length scales beyond {\it{ab initio}} reach, $\approx{5^3}$~nm$^3$--15$^3$~nm$^3$.

Key findings of our work are summarized below.

\subsection*{Methodological: MLIP development}
\begin{enumerate}
    \item MLIPs for simulations of tensile deformation until fracture can be trained following the scheme in Fig.~\ref{FIG: workflow}, based on snapshots from finite-temperature AIMD simulations of sequential elongation with supercells near {\it{ab initio}} size limit ($\approx{10^3}$ atoms). The strategy can be generalized to other loading conditions (e.g. shearing).
    
   \item Due to strain-induced nucleation of extended defects, a transition from atomic to nanoscale simulations---here including also more realistic computational setup (continuous deformation, Poisson's contraction)---may require up-fitting. We propose to generate additional {\it{ab inito}} data by finite-temperature AIMD: imposing a large strain along one lattice vector, initializing atoms at ideal lattice sites, and equilibrating the supercell under fixed volume and shape.
   Again, this can be generalized to, e.g., simple shear or compression.

   \item MLIPs fitted to room-temperature tensile dataset may be transferable to simulate other loading conditions; here we show examples of high-temperature tensile deformation and shear deformation (at the atomic scale). 
   Contrarily, up-fitting is certainly required for simulations of volumetric compression, other phases and stoichiometries.
   
\end{enumerate}

\subsection*{Materials science: predictions for TiB$_2$}
\begin{enumerate}
    \item Our calculations indicate elastic isotropy of TiB$_2$'s basal plane at 300 and 1200\;K. 

    \item Mechanical properties derived from initially dislocation-free nanoscale supercells, with $\approx{10}^4$--10$^6$ atoms, are directionally dependent but well saturated.
    At 300\;K, theoretical tensile strength during [0001], $[10\overline{1}0]$, and $[\overline{1}2\overline{1}0]$ deformation reaches 51--56\;GPa.
    It is the highest during $[\overline{1}2\overline{1}0]$ loading, while toughness is the highest during [0001] loading.

     \item Nanoscale MD simulations provide insights into crack nucleation and growth mechanisms.  
     Subject to [0001] tensile loading, Ti/B$_2$ layer delamination induces opening of nm-sized voids which rapidly coalesce, inducing formation of few-nm-size cracks inside the material.
     Fracture surfaces align predominantly with basal planes, \{0001\}, and first order pyramidal planes, \{10$\overline{1}$1\}. 
     
    \item Considering deformation within the basal plane, $[10\overline{1}0]$ tensile test (i.e. loading in the direction of strong B--B bonds), most often induces crack deflection, formation of V-shaped defects, and fracture at \{11$\overline{2}2\}$ family of surfaces.
    Contrarily, the $[\overline{1}2\overline{1}0]$ tensile deformation induces fracture at \{10$\overline{1}$0\} prismatic planes.

\end{enumerate}

In projection, the here-proposed training strategy should be applicable to other ceramics (particularly transition metal diborides but also e.g. nitrides, carbides) as well as other MLIP formalisms (e.g. neural network potentials).
Our nanoscale simulations for TiB$_2$ provide guidance for interpretation of micromechanics experiments, in particular, can complement post-mortem transmission electron microscopy.
Follow-up work could focus on MLIP up-fitting for more complex loading geometries, e.g., with a pre-crack.

\section*{Methods}
\small

\subsection*{Ab initio molecular dynamics (AIMD)}
Born-Oppenheimer AIMD calculations were carried out using the VASP~\cite{VASP-1} package together with the projector augmented wave (PAW)~\cite{VASP-2} method and the Perdew-Burke-Ernzerhof exchange-correlation functional revised for solids (PBEsol)~\cite{PBEsol}.
All TiB$_2$ supercells were based on the AlB$_2$-type structure (P6/mmm).
The 720-atom (240~Ti$+$480~B) supercell---used to generate the training/learning/validation dataset---had $\approx(1.5 \times 1.6 \times 2.6)~\mathrm{nm}^3$, with $x$, $y$, $z$ axes chosen to satisfy to following crystallographic relationships: $x\parallel[10\overline{1}0]$, $y\parallel[\overline{1}2\overline{1}0]$, $z\parallel[0001]$. 

{\bf{Equilibration}} of TiB$_2$ (at 300 and 1200\;K, with $\Gamma$ point only) was performed in 2 steps: (i) a 10\;ps AIMD run with isobaric-isothermal (NPT) ensemble, Parrinello-Rahman barostat~\cite{NPT} and Langevin thermostat, and (ii) a  2\;ps for 300~K, {4\;ps} for 1200~K AIMD run with the canonical (NVT) ensemble with Nosé-Hoover thermostat, using time-averaged lattice parameters from the NPT simulation (time steps with reasonably converged energy and stresses---see the Suppl. Mat., Fig.~S1).
Computational setup for simulations of [0001], $[10\overline{1}0]$, and $[\overline{1}2\overline{1}0]$ {\bf{tensile deformation}} and (0001)$[\overline{1}2\overline{1}0]$, $(10\overline{1}0)[\overline{1}2\overline{1}0]$, and $(10\overline{1}0)[0001]$ {\bf{shear deformation}} (all with $\Gamma$ point only) followed Refs.~\cite{Method1,Method2,Method3}.
Specifically, the equilibrated supercell was elongated (for shear: tilted) in the desired direction (for shear: angle) with a 2\%-strain step.
No shrinkage of lattice parameters due to Poisson's effect was allowed.
Each deformation step consisted in an 0.3~ps pre-relaxation (isokinetic velocity rescaling) and a 2.7~ps NVT run. 

Stress tensor components were calculated as averages of the last 0.5~ps.
{\bf{Elastic constants}}, $C_{ij}$, were evaluated following Ref.~\cite{sangiovanni2021temperature}, based on a second-order polynomial fit of the [0001], [10$\overline{1}0$], and [$\overline{1}2\overline{1}0$] stress/strain data ($C_{11}$, $C_{12}$, $C_{13}$, $C_{33}$) and of the (0001)$[\overline{1}2\overline{1}0]$, $(10\overline{1}0)[\overline{1}2\overline{1}0]$, and $(10\overline{1}0)[0001]$ shear stress/strain data ($C_{44}$), assuming strains up to 4\%.

For purposes of the Transferability section, simulations of {\bf{volumetric compression}} were carried out for a 720-atom TiB$_2$ supercell maintained at 300\;K (for 2~ps) with the NVT ensemble.
{\bf{Surface energies}} were calculated using a 60-atom TiB$_2$ supercell (${2\times1\times5}$, with a \textbf{k}-mesh of $60$ in each direction) together with a {10\AA} vacuum layer. 
The supercell was fully relaxed at 0\;K (in terms of lattice constants and atomic positions) until forces on atoms were below $10^{-5}$ eV{/\AA} and total energy was converged until 0.01\;eV/supercell. 
Other {\bf{ground-state and higher-energy structures from the Ti--B phase diagram}} (Ti$_2$B, Ti$_3$B$_4$, TiB, TiB$_{12}$, etc.) were fully relaxed at 0\;K starting from lattice parameters and atomic positions from the Materials Project~\cite{MaP}, applying the same convergence criteria as for surface energies.

\subsection*{Development of machine-learning interatomic potentials (MLIPs)}
We used the moment tensor potential (MTP) formalism, as implemented in the MLIP-2 package~\cite{MLIP}.
Training data generation and general workflow are detailed in the section Training dataset generation and workflow. 
Note that the pool of training/learning/validation configurations did not contain snapshots from the very initial stage (first 5\%) of NVT simulations (with potentially largely oscillating total energies). 
MLIPs were fitted based on the 16g MTPs (referring to the highest degree of polynomial-like basis functions in the analytic description of the MTP~\cite{MTP}), using the Broyden-Fletcher-Goldfarb-Shanno method~\cite{fletcher2013practical} with 1500 iterations and 1.0, 0.01 and 0.01 weights for total energy, stresses and forces in the loss functional.
The cutoff radius of 5.5~\AA\ was employed, similar to other recent MLIP studies~\cite{erhard2022machine,Method4} (our tests for larger cutoffs, 7.4 and 10.0~\AA\, did not show notable changes in accuracy).

\subsection*{Molecular dynamics with MLIPs (ML-MD)}
ML-MD calculations were performed with the LAMMPS code~\cite{LAMMPS} interfaced with MLIP-2 package~\cite{MLIP}  which allows to run simulations with MTP-type potentials.
Computational setup of {\bf{atomic scale ML-MD tensile and shear tests}} (at 300 or 1200\;K) was equivalent to AIMD tensile and shear tests described above, in particular, carried out for TiB$_2$ with the same supercell size and orientation. 
Stress tensor components and elastic constants were calculated in the same way as described above in case of AIMD.

{\bf{Nanoscale ML-MD tensile tests}} (at 300 or 1200\;K) used TiB$_2$ supercells with 12,960 atoms ($S1$); 141,120 atoms ($S2$); 230,400 atoms ($S3$); and 432,000 atoms ($S4$); with dimensions of about $(1.5 \times 1.6 \times 2.6)~\mathrm{nm}^3$, $(4.6 \times 4.7 \times 5.1)~\mathrm{nm}^3$, $(10.6 \times 11.0 \times 10.3)~\mathrm{nm}^3$, $(12.1 \times 12.6 \times 12.9)~\mathrm{nm}^3$, and $(15.2 \times 15.8 \times 15.4)~\mathrm{nm}^3$, respectively. 
Prior to simulating mechanical deformation, the supercells were equilibrated for $5$\;ps at the target temperature (300 or 1200\;K) using the isobaric-isothermal (NPT) ensemble coupled to the Nosé-Hoover thermostat with a 1~fs time step.
Tensile loading was simulated by deforming the supercell at each time step with a constant strain rate (50~\AA/s), accounting for lateral contraction (Poisson's effect) via the NPT thermostat.

Atomic scale {\bf{volumetric compression}} simulations used analogical setup and the same supercell sizes as described above in case of AIMD.
{\bf{Surface structures and other Ti--B phases}} were fully relaxed at 0\;K to a local energy minimum using the conjugate gradient method (molecular statics, MS) following analogical setup as described above in case of {\it{ab initio}} calculations.

\subsection*{Visualization and structural analysis}
The OVITO package~\cite{OVITO} allowed to visualize and analyze selected AIMD and ML-MD simulations, in particular, using functions (i) {\it{Radial pair distribution function}} (with a cut-off radius of 5.5~\AA, i.e. the cutoff of our potentials), (ii) {\it{Elastic strain calculation}} and (iii) {\it{Atomic strain}} (with cut-off radius of $\pm$~0.1~mm).
For details see the OVITO documentation.

\bibliography{references}

\section*{Acknowledgements}
\textcolor{red}{SL acknowledges xx.}
\textcolor{red}{LCT acknowledges xx.}
\textcolor{red}{FT acknowledges xx.}
LH acknowledges financial support from the Swedish Government Strategic Research Area in Materials Science on Functional Materials at Linköping University SFO-Mat-LiU No. 2009 00971.
Support from Knut and Alice Wallenberg Foundation Scholar Grants KAW2016.0358 and KAW2019.0290 is also acknowledged by LH.
\textcolor{red}{PHM acknowledge xx.}
DGS gratefully acknowledges financial support from the Swedish Research Council (VR) through Grant Nº VR-2021-04426 and the Competence Center Functional Nanoscale Materials (FunMat-II) (Vinnova Grant No. 2022-03071).
NK acknowledges the Austrian Science Fund, FWF, (T-1308). 
The computations handling were enabled by resources provided by the National Academic Infrastructure for Supercomputing in Sweden (NAISS) and the Swedish National Infrastructure for Computing (SNIC) at Sigma and Tetralith Clusters partially funded by the Swedish Research Council through grant agreements no. 2022-06725 and no.~2018-05973, as well as by the Vienna Scientific Cluster (VSC) in Austria. 
The authors acknowledge TU Wien Bibliothek for financial support through its Open Access Funding Program.

\section*{Author contributions statement}
NK conceived and designed the project; SL carried out the simulations, analyzed the data and wrote the manuscript; NK and DGS advised on the calculations and manuscript writing; 
LCT and FT gave constructive suggestions during the whole process and modified the manuscript;
LH and PHM supported the project financially and modified the manuscript. All authors discussed the results and commented on the manuscript.

\section*{Conflicts of interest}
The authors declare that they have no known competing financial interests or personal relationships that could have appeared to influence the work reported in this paper.

\end{document}



\section*{Machine-learning potentials for nanoscale simulations of deformation and fracture: example of TiB$_2$ ceramic}

{\large ${  }^{ }$ Shuyao Lin ${ }^{1,2^*}$, Luis Casillas-Trujillo$^2$, Ferenc Tasnádi$^2$, Lars Hultman ${ }^2$, Paul H. Mayrhofer ${ }^1$, Davide G. Sangiovanni ${ }^2$, and Nikola Koutná ${ }^{1,2}$}

${ }^1$ Technischen Universität Wien, Institute of Materials Science and Technology, Vienna, A-1060, Austria 

${ }^2$ Linkōping University, Department of Physics, Chemistry, and Biology (IFM), Linkōping, SE-58183, Sweden 

$*$ shuyao.lin@tuwien.ac.at

\section*{Supplementary Materials}

\begin{figure}[H]
\centering
\includegraphics[width=1.0\textwidth]{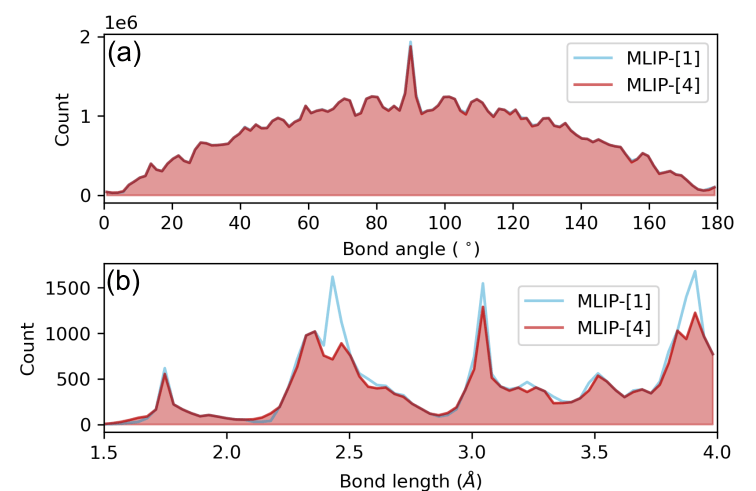}
\caption{Comparison of (a) bond angles and (b) bond lengths in TiB$_2$ configurations obtained during MD tensile testing based on MLIP-[1] and MLIP-[4] potentials.}
\label{SupplFIG: RDF-Bond}
\end{figure}

\begin{itemize}
\item {\bf{Finite-temperature equilibrium lattice parameters of TiB$_2$ supercells.}}

     Fig.~\ref{SupplFIG: SR} illustrates our procedure used to find equilibrium lattice parameters of a 720-atom TiB$_2$ supercell at 300 and 1200\;K, which is later employed in AIMD and ML-MD tensile tests (Section ``Training procedure and fitting initial MLIPs'' in the manuscript). Specifically, Fig.~\ref{SupplFIG: SR}(a-b) depict volume, total energy, and temperature evolution during AIMD simulations with the NPT ensemble. Equilibration is reached after 5 ps: steady trends in property vs time. Thus, equilibrium lattice parameters are calculated by time-averaging a, b, and c after 5 ps.
     Fig.~\ref{SupplFIG: SR}(c-d) depict total energy evolution during a subsequent 2~ps NVT simulations with the above time-averaged lattice parameters.

\item {\bf{Validation of ML potentials MLIP-[1{-}4] against room-temperature (MLIP-[1-3]) and high-temperature (MLIP-[4]) AIMD tensile tests.}}

    Fig.~\ref{SupplFIG: MTP-1to3}--\ref{SupplFIG: MTP-4} depict comparison of AIMD and ML-MD stress strain curves and snapshots of fracture points during [0001], [10$\overline{1}0$], and [$\overline{1}2\overline{1}0$] tensile deformation of TiB$_2$, using a 720-atom supercell.
    Specifically, Fig.\ref{SupplFIG: MTP-1to3}a shows that each trained MLIP yields sufficient agreement with AIMD results of atomic scale tensile tests along all the three directions. Fig.\ref{SupplFIG: MTP-1to3}c-d show that the fracture pathways observed in AIMD simulations are correctly reproduced (see Fig.\ref{SupplFIG: MTP-1to3}b).
    The behavior of MLIP-[4] during atomic-scale simulations at 300~K is discussed and validated in the main manuscript. Fig.\ref{SupplFIG: MTP-4} illustrates very good agreement between ML-MD and AIMD calculations for both stress-strain curves and fracture mechanisms at 1200~K.

\item {\bf{Performance of MLIP-[1--4] for nanoscale ML-MD simulations.}}

   Fig.\ref{SupplFIG:Poisson}a,b exhibits the fracture behavior of nanoscale-S1 along [0001] direction, as predicted by MLIP-[1] (a) and MLIP-[2,3] (b) during MD tensile tests. These potentials produce similar supercell behavior for stresses approaching the tensile strength.
   The similarity of the results obtained with MLIP-[2,3] indicates that the inclusion of high-temperature strained configurations to the training set is not of critical relevance for obtaining realistic fracture mechanisms.
   Fig.\ref{SupplFIG:Poisson}c, instead, shows that the addition of highly-deformed structures and fracture surfaces to the training set enables smooth continuous elongation and describes fracture pathways in a reasonable manner.

   Fig.\ref{SupplFIG: RDF-B1} gives a comparison between the radial distriburion function of different bonding in the nanoscale (S1) TiB$_{2}$ during tensile tests implemented with MLIP-[4] along [0001] (a), $[10\overline{1}0]$ (b) and $[\overline{1}2\overline{1}0]$(c). In each panel, curves of different color correspond to different deformation stages: initial step (blue), one step before fracture (orange), during fracture (green and purple), after fracture (red and brown).
   Fig.\ref{SupplFIG: RDF-Bond} presents geometrical differences between chemical environments produced by MD tensile tests based on MLIP-[1] and MLIP-[4]. From the analysis, one can conclude that there is no significant difference between bond angles and bond lengths in configurations obtained with these two potentials. The main difference comes from
configurations with multiple energy and stress mapping (\textcolor{violet}{@Nikola: from high-energy configurations?}), as discussed in the main text under the ``MLIPs' up-fitting for nanoscale tensile tests'' section.

\end{itemize}

\begin{figure}[H]
\centering
\includegraphics[width=1.0\textwidth]{Figures/Sup-1-NPT_NVT-Mod-M.png}
\caption{Volume, total energy, and temperature evolution during NPT simulations with a 720-atom TiB$_2$ supercell at (a) 300 and (b) 1200\;K.
Total energy variation during a subsequent NVT simulation with time-averaged lattice parameters---calculated based on the NPT data after the red vertical line (satisfactory convergence of volume, total energy, and temperature)--- at (c) 300~K and (d) 1200~K.
}
\label{SupplFIG: SR}
\end{figure}

\begin{figure}[H]
\centering
\includegraphics[width=1.0\textwidth]{Figures/Sup-Val-MTP-1to3-Full-M.png}
\caption{Performance of ML potentials MLIP-[1,2,3].
    (a) Comparison of AIMD and ML-MD stress/strain curves for TiB$_2$ subject to $[0001]$, $[10\overline{1}0]$, and $[\overline{1}2\overline{1}0]$ tensile deformation at 300~K, using a 720-atom supercell.
    Only the stress component in the loaded direction is plotted.
    (b) Snapshots of the AIMD fracture points.
    (c) Snapshots of the ML-MD (through MLIP-[1]) fracture points.
    (d) Snapshots of the ML-MD (through MLIP-[2]) fracture points.
    (e) Snapshots of the ML-MD (through MLIP-[3]) fracture points.
    }
\label{SupplFIG: MTP-1to3}
\end{figure}

\begin{figure}[H]
\centering
\includegraphics[width=0.9\textwidth]{Figures/Sup-Val-MTP-4-HighT-Screenshot.png}
\caption{Performance of ML potential MLIP-[4].
    (a) Comparison of AIMD and ML-MD stress/strain curves for TiB$_2$ subject to $[0001]$, $[10\overline{1}0]$, and $[\overline{1}2\overline{1}0]$ tensile deformation at 1200~K, using a 720-atom supercell.
    Only the stress component in the loaded direction is plotted.
    (b) Snapshots of the AIMD fracture points.
    (c) Snapshots of the ML-MD fracture points.}
\label{SupplFIG: MTP-4}
\end{figure}

\begin{figure}[H]
\centering
\includegraphics[width=0.9\textwidth]{Figures/Sup-3-4-SingleTi-Other-Poisson-M.png}
\caption{MD nanoscale-scale simulations of $S1$ cell strained along [0001] implemented with (a)~MLIP-[1], (b)~MLIP-[2,3], (c)~MLIP-[4]. The simulations implemented with MLIP-[1-3] are valid only up to strains approaching maximum elongation. Contrarily, nanoscale simulations based on MLIP-[4] can realistically describe fracture.
}
\label{SupplFIG:Poisson}
\end{figure}

\begin{figure}[H]
\centering
\includegraphics[width=1.0\textwidth]{Figures/Main-3-B1-SingleTi-Coordination-Analysis.png}
\caption{Radial distribution function (RDF, with 5.5~\AA\ cutoff) for Ti--Ti, Ti--B, and B--B bonds (integrated over all configurations) of S1-cell MD tensile tests with MLIP-[4] along (a)~$[0001]$, (b)~$[10\overline{1}0]$, and (c)~$[\overline{1}2\overline{1}0]$. Curves of different colors indicate different tensile steps: initial step (blue and 0$\%$), one step before fracture (orange), during fracture (green and purple), after fracture (red and brown). The purple and brown curves represent only fracture regions, the others represent the whole configuration.}
\label{SupplFIG: RDF-B1}
\end{figure}

